\documentclass[preprint,superscriptaddress,nofootinbib]{revtex4}

     \usepackage{amssymb,amsmath,graphicx,subfigure,color}
     \usepackage{rotating,multirow,dcolumn,float}
     \usepackage[colorlinks,breaklinks]{hyperref}
     \usepackage{orcidlink}
     \allowdisplaybreaks[4]
     \begin{document}

     \title{Phenomenological revisit of the $B$ ${\to}$ ${\pi}{\pi}$ decays \\
            by incorporating soft color-octet contributions}
     \author{Yueling Yang \orcidlink{0000-0002-7688-2076}}
     \affiliation{School of Physics,
                 Henan Normal University, Xinxiang 453007, China}
     \author{Bingbing Yang \orcidlink{0009-0004-0575-6955}}
     \affiliation{School of Physics,
                 Henan Normal University, Xinxiang 453007, China}
     \author{Junfeng Sun \orcidlink{0000-0003-4742-4292}}
     \affiliation{School of Physics,
                 Henan Normal University, Xinxiang 453007, China}

     \begin{abstract}
     Motivated by the  long-standing ``${\pi}{\pi}$ puzzle'' in
     nonleptonic $B$ decays, we revisit the $B$ ${\to}$
     ${\pi}{\pi}$ decays within the QCD factorization (QCDF)
     framework, now incorporating the soft color-octet matrix
     element (COME) contributions, an ingredient that has been
     largely overlooked in earlier QCDF studies.
     A comprehensive phenomenological analysis is performed
     under three distinct scenarios, with the free parameters
     determined via fits to the available experimental data
     using the ${\chi}^{2}$ method.
     The goodness-of-fit is very poor for Scenario S1, which
     corresponds to the traditional QCDF case without the
     soft COME contributions,
     and for Scenario S2, which includes the soft COME contributions
     but omits the weak annihilation (WA) contributions.
     Scenario S3, in which both the COME and WA contributions
     are included simultaneously, achieves an excellent overall
     fit. In Scenario S3, the QCDF approach can
     accommodate almost all $B$ ${\to}$ ${\pi}{\pi}$
     observables in a consistent and satisfactory manner, and
     the fitted parameters ${\rho}_{A}$ $<$ $1$ (WA) and
     ${\rho}_{c}$ ${\sim}$ $0.3$ (soft COME) are in good agreement
     with the QCDF power counting scheme and the factorization
     hypothesis.

     \href{http://arxiv.org/abs/2502.12461}
          {arXiv:2502.12461}

     \href{https://doi.org/10.1103/wrm3-6qwv}
          {Phys. Rev. D 114, 096006 (2026)} ~~
     \href{https://doi.org/10.1103/wrm3-6qwv}
          {https://doi.org/10.1103/wrm3-6qwv}
     \end{abstract}
     \maketitle

     \section{Introduction}
     \label{sec01}
     From the perspective of the underlying weak interaction dynamics,
     the $B$ ${\to}$ ${\pi}{\pi}$ decays are effectively induced by
     the quark transition $b$ ${\to}$ $u\bar{u}d$ within the
     Standard Model (SM) of elementary particle physics.
     These decays serve as significant and fascinating probes
     for investigating the origin of $CP$ violation in weak decays
     of flavored hadrons, and for constraining the angle ${\alpha}$
     sandwiched between the sides $ V_{ud} V_{ub}^{\ast} $ and
     $ V_{td} V_{tb}^{\ast} $ of the most commonly studied unitarity
     triangle
     \cite{PDG.2026,PhysRevD.41.1522,PhysRevLett.65.3381,PhysRevD.44.1454},
     where $V_{ij}$ denotes the Cabibbo–Kobayashi–Maskawa (CKM) matrix
     elements \cite{PRL.10.531,PTP.49.652}.
     Moreover, the two-body kinematics of the $B$ ${\to}$ ${\pi}{\pi}$
     decays are remarkably simple, which facilitates a detailed
     investigation of weak decay dynamics and the behavior of
     QCD 
     across different energy regimes,
     while also providing a foundation for studies of multi-body
     $B$ meson decays.

     \begin{table}[b]
     \caption{Experimental data on branching ratios and $CP$
        asymmetries for the $B$ ${\to}$ ${\pi}{\pi}$ decays.}
     \label{tab:exp-data}
     \begin{ruledtabular}
     \begin{tabular}{ccccc}
     & PDG \cite{PDG.2026}
     & Belle
     & BaBar
     & Belle II
       \\ \hline
       \begin{math} 10^{6} \, {\times} \, {\cal B}( {\pi}^{-} {\pi}^{0} ) \end{math}
     & \begin{math} 5.31 {\pm} 0.26 \end{math}
     & \begin{math} 5.86 {\pm} 0.46 \end{math} \cite{PhysRevD.87.031103}
     & \begin{math} 5.02 {\pm} 0.54 \end{math} \cite{PhysRevD.76.091102}
     & \begin{math} 5.10 {\pm} 0.40 \end{math} \cite{PhysRevD.109.012001}
       \\
       \begin{math} 10^{6} \, {\times} \, {\cal B}( {\pi}^{0} {\pi}^{0} ) \end{math}
     & \begin{math} 1.46 {\pm} 0.19 \end{math}
     & \begin{math} 1.31 {\pm} 0.27 \end{math} \cite{PhysRevD.96.032007}
     & \begin{math} 1.83 {\pm} 0.25 \end{math} \cite{PhysRevD.87.052009}
     & \begin{math} 1.25 {\pm} 0.23 \end{math} \cite{PhysRevD.111.L071102}
       \\
       \begin{math} 10^{6} \, {\times} \, {\cal B}( {\pi}^{+} {\pi}^{-} ) \end{math}
     & \begin{math} 5.43 {\pm} 0.26 \end{math}
     & \begin{math} 5.04 {\pm} 0.28 \end{math} \cite{PhysRevD.87.031103}
     & \begin{math} 5.5~ {\pm} 0.5~ \end{math} \cite{PhysRevD.75.012008}
     & \begin{math} 5.83 {\pm} 0.28 \end{math} \cite{PhysRevD.109.012001}
       \\ \hline
       \begin{math} {\cal A}_{CP}( {\pi}^{-} {\pi}^{0} ) \end{math}
     & \begin{math} - 0.01~ {\pm} 0.04~ \end{math}
     & \begin{math}   0.025 {\pm} 0.044 \end{math} \cite{PhysRevD.87.031103}
     & \begin{math} ~ 0.03~ {\pm} 0.08~ \end{math} \cite{PhysRevD.76.091102}
     & \begin{math} - 0.081 {\pm} 0.055 \end{math} \cite{PhysRevD.109.012001}
       \\
       \begin{math} {\cal C}_{CP}( {\pi}^{0} {\pi}^{0} ) \end{math}
     & \begin{math} - 0.23~ {\pm} 0.18~ \end{math}
     & \begin{math} - 0.14  {\pm} 0.37~ \end{math} \cite{PhysRevD.96.032007}
     & \begin{math} - 0.43~ {\pm} 0.26~ \end{math} \cite{PhysRevD.87.052009}
     & \begin{math} ~ 0.05~ {\pm} 0.29~ \end{math} \cite{2606.20797}
       \\
       \begin{math} {\cal S}_{CP}( {\pi}^{0} {\pi}^{0} ) \end{math}
     & & & & \begin{math} ~ 0.61~ {\pm} 0.80~ \end{math} \cite{2606.20797}
       \\
       \begin{math} {\cal C}_{CP}( {\pi}^{+} {\pi}^{-} ) \end{math}
     & \begin{math} - 0.314 {\pm} 0.030 \end{math}
     & \begin{math} - 0.33  {\pm} 0.07~ \end{math} \cite{PhysRevD.88.092003}
     & \begin{math} - 0.25~ {\pm} 0.08~ \end{math} \cite{PhysRevD.87.052009}
     &
       \\
       \begin{math} {\cal S}_{CP}( {\pi}^{+} {\pi}^{-} ) \end{math}
     & \begin{math} - 0.670 {\pm} 0.030 \end{math}
     & \begin{math} - 0.64  {\pm} 0.09~ \end{math} \cite{PhysRevD.88.092003}
     & \begin{math} - 0.68~ {\pm} 0.10~ \end{math} \cite{PhysRevD.87.052009}
     &
     \end{tabular}
     \end{ruledtabular}
     \end{table}

     Over the past years, the $B$ ${\to}$ ${\pi}{\pi}$ decays have
     attracted considerable interest from both experimental and
     theoretical communities, leading to numerous significant
     advances.
     On the experimental side, a wealth of data on $B$ meson decays
     is now available and continues to accumulate with steadily
     improving precision.
     The most recent experimental results for the branching ratios
     and $CP$ asymmetries of the $B$ ${\to}$ ${\pi}{\pi}$ decays
     are summarized in Table~\ref{tab:exp-data}.
     On the theoretical side, several QCD-improved factorization
     approaches have been developed to compute hadronic matrix
     elements (HMEs) effectively,
     including the QCD factorization (QCDF) approach
     \cite{PhysRevLett.83.1914,NuclPhysB.591.313,NuclPhysB.606.245,
     PhysLettB.488.46,PhysLettB.509.263,PhysRevD.64.014036}
     and the perturbative QCD (PQCD) approach
     \cite{PhysRevLett.74.4388,PhysLettB.348.597,PhysLettB.504.6,
     PhysRevD.63.054008,PhysRevD.63.074006,PhysRevD.63.074009,
     EPJC.23.275,PhysLettB.555.197}.
     Selected branching ratios predicted by these phenomenological
     approaches are listed in Table~\ref{tab:theo-br}.

     \begin{table}[b]
     \caption{Theoretical results for branching ratios of the $B$
       ${\to}$ ${\pi}{\pi}$ decays.
       For further details, see the references cited.}
     \label{tab:theo-br}
     \begin{ruledtabular}
     \begin{tabular}{c|l ccc}
       \multicolumn{2}{c}{cases}
     & \begin{math} 10^{6} \, {\times} \, {\cal B}( {\pi}^{-} {\pi}^{0} ) \end{math}
     & \begin{math} 10^{6} \, {\times} \, {\cal B}( {\pi}^{0} {\pi}^{0} ) \end{math}
     & \begin{math} 10^{6} \, {\times} \, {\cal B}( {\pi}^{+} {\pi}^{-} ) \end{math}
     \\ \hline
     & Ref. \cite{PhysRevD.65.074001}
     & $ 5.36 $
     & $ 0.13 $
     & $ 9.97 $  \\
     & Ref. \cite{NuclPhysB.675.333}\footnotemark[1]
     & $ 6.0^{ +3.0 +2.1 +1.0 +0.4 }_{ -2.4 -1.8 -0.5 -0.4 } $
     & $ 0.3^{ +0.2 +0.2 +0.3 +0.2 }_{ -0.2 -0.1 -0.1 -0.1 } $
     & $ 8.9^{ +4.0 +3.6 +0.6 +1.2 }_{ -3.4 -3.0 -1.0 -0.8 } $  \\
    && ( $ 5.8 $, $ 5.5 $, $ 6.0 $, $ 5.1 $ )
     & ( $ 0.7 $, $ 0.9 $, $ 0.4 $, $ 0.7 $ )
     & ( $ 6.0 $, $ 4.6 $, $ 9.5 $, $ 5.2 $ ) \\
     & Ref. \cite{NuclPhysB.832.109}\footnotemark[2]
     &   $ 5.43^{ +0.06 +1.45 }_{ -0.06 -0.84 } $
     &   $ 0.33^{ +0.11 +0.42 }_{ -0.08 -0.17 } $
     &   $ 7.37^{ +0.86 +1.22 }_{ -0.69 -0.97 } $   \\
    && ( $ 5.82^{ +0.07 +1.42 }_{ -0.06 -1.35 } $ )
     & ( $ 0.63^{ +0.12 +0.64 }_{ -0.10 -0.42 } $ )
     & ( $ 5.70^{ +0.70 +1.16 }_{ -0.55 -0.97 } $ ) \\
       QCDF
     & Ref. \cite{PLB.748.422}\footnotemark[3]
     &   $ 5.26^{ +1.13 }_{ -1.04 } $
     &   $ 0.39^{ +0.09 }_{ -0.09 } $
     &   $ 5.93^{ +1.67 }_{ -1.47 } $   \\
    && ( $ 5.89^{ +1.25 }_{ -1.23 } $ )
     & ( $ 0.98^{ +0.44 }_{ -0.31 } $ )
     & ( $ 5.60^{ +1.19 }_{ -1.68 } $ ) \\
     & Ref. \cite{PhysRevD.80.114008}
     & $ 5.9^{ +2.2 +1.4 }_{ -1.1 -1.1 } $
     & $ 1.1^{ +1.0 +0.7 }_{ -0.4 -0.3 } $
     & $ 7.0^{ +0.4 +0.7 }_{ -0.7 -0.7 } $ \\
     & Ref. \cite{PhysRevD.90.054019}\footnotemark[4]
     & ( $ 5.20^{ +0.64 +1.11 }_{ -0.47 -1.00 } $ )
     & ( $ 1.67^{ +0.22 +0.25 }_{ -0.19 -0.23 } $ )
     & ( $ 5.88^{ +0.66 +1.66 }_{ -0.49 -1.45 } $ )  \\
    && [ $ 5.59^{ +0.68 +1.15 }_{ -0.51 -1.04 } $ ]
     & [ $ 2.13^{ +0.29 +0.32 }_{ -0.24 -0.29 } $ ]
     & [ $ 5.74^{ +0.64 +1.63 }_{ -0.47 -1.42 } $ ]  \\ \hline
       SCET
     & Ref. \cite{PhysRevD.74.014003}
     & $ 5.2  {\pm} 1.6  {\pm} 2.1  {\pm} 0.6  $
     & $ 0.84 {\pm} 0.29 {\pm} 0.30 {\pm} 0.19 $
     & $ 5.4  {\pm} 1.3  {\pm} 1.4  {\pm} 0.4  $ \\ \hline
     & Ref. \cite{PhysRevD.90.014029}
     & $ 4.27^{ +1.85 }_{ -1.47 } $
     & $ 0.23^{ +0.19 }_{ -0.15 } $
     & $ 7.67^{ +3.47 }_{ -2.64 } $ \\
     & Ref. \cite{PhysRevD.91.114019}\footnotemark[5]
     &   $ 3.35^{ +1.08 +0.23 }_{ -0.77 -0.22 } $
     &   $ 0.29^{ +0.11 +0.03 }_{ -0.07 -0.02 } $
     &   $ 6.19^{ +2.09 +0.38 }_{ -1.48 -0.34 } $   \\
    && ( $ 4.45^{ +1.38 +0.39 }_{ -0.99 -0.36 } $ )
     & ( $ 0.61^{ +0.16 +0.14 }_{ -0.12 -0.12 } $ )
     & ( $ 5.39^{ +1.86 +0.28 }_{ -1.31 -0.25 } $ ) \\
     & Ref. \cite{PhysRevD.103.056006}\footnotemark[6]
     &  ( $ 3.04^{ +0.18 +0.16 +0.27 }_{ -0.17 -0.16 -0.25} $ )
     &  ( $ 0.24^{ +0.02 +0.04 +0.01 }_{ -0.02 -0.03 -0.01} $ )
     &  ( $ 5.52^{ +0.34 +0.37 +0.47 }_{ -0.32 -0.36 -0.43} $ ) \\  PQCD &
     &  [ $ 3.23^{ +0.19 +0.18 +0.29 }_{ -0.18 -0.18 -0.27} $ ]
     &  [ $ 0.24^{ +0.02 +0.04 +0.01 }_{ -0.02 -0.04 -0.01} $ ]
     &  [ $ 5.82^{ +0.36 +0.41 +0.50 }_{ -0.34 -0.39 -0.46} $ ] \\
    && \{ $ 3.42^{ +0.21 +0.20 +0.32 }_{ -0.19 -0.20 -0.30} $ \}
     & \{ $ 0.23^{ +0.02 +0.04 +0.01 }_{ -0.02 -0.04 -0.01} $ \}
     & \{ $ 6.12^{ +0.39 +0.45 +0.54 }_{ -0.36 -0.43 -0.50} $ \} \\
     & Ref. \cite{ChinPhysC.46.123103}
     &   $ 4.18^{ +1.32 }_{ -0.97 } $
     &   $ 0.23^{ +0.07 }_{ -0.05 } $
     &   $ 7.31^{ +2.38 }_{ -1.72 } $   \\
     & Ref. \cite{PhysRevD.107.013004}
     &   $ 5.72 {\pm} 0.44 ^{ +0.29 }_{ -0.37 } $
     &   $ 1.50 {\pm} 0.24 ^{ +0.18 }_{ -0.19 } $
     &   $ 5.14 {\pm} 0.61 ^{ +0.34 }_{ -0.37 } $   \\
     & Ref. \cite{PhysRevD.110.056025}
     &   $ 5.63 {\pm} 0.53 ^{ +0.12 +0.19 }_{ -0.23 -0.20 } $
     &   $ 1.41 {\pm} 0.35 ^{ +0.04 +0.12 }_{ -0.06 -0.17 } $
     &   $ 5.14 {\pm} 0.69 ^{ +0.16 +0.28 }_{ -0.25 -0.24 } $
     \end{tabular}
     \end{ruledtabular}
     \end{table}
     \footnotetext[1]{Numbers in parentheses correspond to four different scenarios.}
     \footnotetext[2]{The NNLO vertex corrections are considered.
                      Numbers in parentheses correspond to the favored results.}
     \footnotetext[3]{Numbers in parentheses are obtained by applying the principle of maximum conformality.}
     \footnotetext[4]{Numbers in round and square brackets correspond to the S1 and S2 scenarios, respectively.}
     \footnotetext[5]{Numbers in parentheses include the Glauber-gluon effect.}
     \footnotetext[6]{Numbers in round, square and curly brackets correspond to the S1, S2 and S3 scenarios, respectively.}

     A comparison of Tables~\ref{tab:exp-data} and \ref{tab:theo-br}
     reveals certain discrepancies between theoretical expectations
     and experimental measurements of branching ratios.
     These discrepancies highlight a long-standing unresolved puzzle
     in nonleptonic $B$ meson decays: how to account simultaneously
     and consistently for all the $B$ ${\to}$ ${\pi}{\pi}$ observables
     within the SM,
     a problem commonly referred to as the ``${\pi}{\pi}$ puzzle''.

     A closer examination of the ``${\pi}{\pi}$ puzzle'' reveals
     that the most conspicuous discrepancy lies, in most cases,
     in the branching ratio for the color-suppressed
     $B$ ${\to}$ ${\pi}^{0}{\pi}^{0}$ decay.
     As shown in Table~\ref{tab:theo-br}, the inclusion of higher
     order QCD radiative corrections, extending in some cases to
     the NNLO level \cite{NuclPhysB.832.109}, still proves
     insufficient to reconcile with the experimental data.
     The theoretical predictions obtained with the
     QCDF \cite{PhysRevD.65.074001,PLB.748.422,
     NuclPhysB.675.333,NuclPhysB.832.109},
     SCET \cite{PhysRevD.74.014003}, and
     PQCD \cite{PhysRevD.90.014029,PhysRevD.91.114019,
     PhysRevD.103.056006,ChinPhysC.46.123103} approaches
     lie significantly below the measured branching ratios.
     This observation suggests that the discrepancy may originate,
     at least in part, from overlooked physical mechanisms beyond
     the perturbative framework.
     Scenario S1 of the data-driven fit in Ref. \cite{PhysRevD.90.054019}
     is conceptually appealing; however, it requires a relatively
     large annihilation parameter ${\rho}_{A}$ ${\sim}$ $3$,
     which appears implausible within the QCDF approach.
     Alternatively, by introducing nonperturbative contributions from
     soft form factors and soft color-octet matrix elements (COME),
     the improved PQCD calculations in
     Refs. \cite{PhysRevD.107.013004,PhysRevD.110.056025}
     achieve a more favorable comparison with the data.
     Nevertheless, an inverted hierarchical branching ratio relation,
     ${\cal B}({\pi}^{+} {\pi}^{-})$ $<$  ${\cal B}({\pi}^{-} {\pi}^{0})$,
     is predicted, which is inconsistent with the measurements
     reported by BaBar, Belle II, and the PDG groups,
     as well as with the majority of theoretical expectations
     (see Tables~\ref{tab:exp-data} and \ref{tab:theo-br}).
     The method/ansatz introduced in Ref. \cite{PhysRevD.107.013004}
     has also been straightforwardly generalized to other nonleptonic
     two-body $B$ meson decays within the modified PQCD approach,
     achieving improved agreement between theoretical calculations
     and experimental measurements
     \cite{PhysRevD.110.056025,PhysRevD.108.013003,PhysRevD.111.036007,
     EPJC.85.146,EPJC.85.821,JPG.52.105001}.
     It is conceivable that some of the insights gleaned from
     these investigations could be profitably employed to refine
     the QCDF calculations.

     It is worth noting that COME contributions are closely
     related to the rationale for the conventional factorization
     approximation and to the dynamical understanding of color
     selection rules in exclusive nonleptonic meson weak decays.
     These contributions have been extensively discussed and
     investigated in connection with apparent discrepancies
     between theoretical predictions and experimental results
     in numerous works, including Refs.
     \cite{PhysRevD.107.013004,PhysRevD.110.056025,
     PhysRevD.108.013003,PhysRevD.111.036007,
     EPJC.85.146,EPJC.85.821,JPG.52.105001,
     PLB.52.351,PLB.86.343,PLB.90.431,PLB.93.125,NPB.171.413,
     NPB.181.74,NPB.183.367,PLB.100.420,PLB.109.227,PhysRevD.25.178,
     NPB.268.16,ZPC.32.237,IJMPA.4.495,PLB.318.549,NPB.389.534,
     NPB.399.441,PLB.335.428,PLB.349.548,PhysRevD.51.3518,
     PhysRevD.51.6259,PhysRevD.53.2506,ZPC.69.253,ZPC.69.647,
     ZPC.71.101,ZPC.72.91,PLB.393.149,PLB.395.345,ZPC.75.651,
     ZPC.76.311,JPG.23.807,EPJC.4.669,PhysRevD.58.094005,
     Adv.Ser.Direct.HEP.15.294,PhysRevD.59.054020,MPLA.14.1841,
     PhysRevD.59.074003,PhysRevD.59.092004,PhysRevD.60.114019,
     NPB.605.558,PLB.571.75,PhysRevD.68.114009,PhysRevD.72.094012,
     JPG.34.637,PLB.675.196,
     CPC.46.073105,JHEP.2023.10.180,CPC.49.023105}.
     In this paper, we will revisit and reassess the soft
     COME contributions as a means of reinvestigating the
     $B$ ${\to}$ ${\pi}{\pi}$ decays within the QCDF framework.
     Our objectives are twofold: first, to ascertain whether
     this ingredient can alleviate the tensions inherent in
     the ``${\pi}{\pi}$ puzzle'', and second,
     to explore the extent to which soft COME contributions are
     empirically supported by the available data.

     The remainder of this paper is organized as follows.
     Section \ref{sec02} delineates the theoretical framework for
     the $B$ ${\to}$ ${\pi}{\pi}$ decays.
     The numerical results and discussion are presented in Section \ref{sec03}.
     A brief summary is given in Section \ref{sec04}.

     \section{Theoretical framework for the $B$ ${\to}$ ${\pi}{\pi}$ decays}
     \label{sec02}
     \subsection{The effective Hamiltonian}
     \label{sec0201}
     By means of the operator product expansion and
     the renormalization group (RG) techniques,
     the effective Hamiltonian governing the nonleptonic
     $B$ ${\to}$ ${\pi}{\pi}$ decays can be expressed as a sum of
     local four-quark operators $O_{i}$, each multiplied by the
     corresponding Wilson coefficients $C_{i}$ and products
     of the CKM matrix elements $V_{ij}$.
     More specifically, this effective Hamiltonian takes the form
     \cite{RevModPhys.68.1125}:
     \begin{equation}
    {\cal H}_{\rm eff} \, = \,
     \frac{G_{F}}{\sqrt{2}} \,
       \sum\limits_{p=u,c} V_{pb} \, V_{pd}^{\ast} \, \Big\{
       \sum\limits_{i=1}^{2} \, C_{i}({\mu}) \, O^{p}_{i}
     + \sum\limits_{j=3}^{10} \, C_{j}({\mu}) \, O_{j} \Big\}
     + {\rm h.c.}
     \label{hamilton},
     \end{equation}
     where $G_{F}$ ${\approx}$ $1.166 {\times} 10^{-5}$ ${\rm GeV}^{-2}$
     \cite{PDG.2026} is the Fermi coupling constant.
     Employing the Wolfenstein parametrization \cite{PRL.51.1945,PRD.50.3433},
     the relevant CKM factors are expressed,
     up to order ${\cal O}({\lambda}^{7})$, as follows:
     \begin{equation}
     V_{ub} \, V_{ud}^{\ast} \, = \,
     A \, {\lambda}^{3} \, ( \bar{\rho} - i \, \bar{\eta} )
     + {\cal O}({\lambda}^{7})
     \label{eq:vub-vud},
     \end{equation}
     \begin{equation}
     V_{cb} \, V_{cd}^{\ast} \, = \, - A \, {\lambda}^{3}
     + {\cal O}({\lambda}^{7})
     \label{eq:vcb-vcd},
     \end{equation}
     \begin{equation}
     V_{tb} \, V_{td}^{\ast} \, = \,
     A \, {\lambda}^{3} \, ( 1 - \bar{\rho} + i \, \bar{\eta} )
     + {\cal O}({\lambda}^{7})
     \label{eq:vtb-vtd}.
     \end{equation}
     The global fit results for the Wolfenstein parameters
     $A$, ${\lambda}$, $\bar{\rho}$, and $\bar{\eta}$ are
     displayed in Table~\ref{tab:input}.
     The Wilson coefficients $C_{i}$ encode the short-distance
     physics contributions arising from gluon exchanges with
     virtualities above the scale ${\mu}$, and are calculated
     using RG-improved perturbation theory,
     owing to the asymptotic freedom of QCD.
     Their numerical values, including both the leading-order (LO) and
     next-to-leading-order (NLO) corrections within the naive dimensional
     regularization (NDR) scheme, are shown in Table~\ref{tab:wcci}.
     The relevant operators governing the decays under consideration
     are expressed as follows:
      \begin{eqnarray}
      O^{p}_{1} & = &
      ( \bar{p}_{\alpha} \, b_{\alpha} )_{V-A} \,
      ( \bar{d}_{\beta}  \, p_{\beta}  )_{V-A}
      \label{operator-01}, \\
      O^{p}_{2} & = &
      ( \bar{p}_{\alpha} \, b_{\beta}  )_{V-A}  \,
      ( \bar{d}_{\beta}  \, p_{\alpha} )_{V-A}
      \label{operator-02}, \\
      O_{i=3,9} & = &
       (\bar{d}_{\alpha} \, b_{\alpha})_{V-A} \, \sum\limits_{q} \,
        \big( {\delta}_{i,3} + {\delta}_{i,9} \, \frac{3}{2} \, Q_{q} \big) \,
       (\bar{q}_{\beta}  \, q_{\beta} )_{V-A}
      \label{operator-03}, \\
      O_{i=4,10} & = &
       (\bar{d}_{\alpha} \, b_{\beta})_{V-A} \, \sum\limits_{q} \,
        \big( {\delta}_{i,4} + {\delta}_{i,10} \, \frac{3}{2} \, Q_{q} \big) \,
       (\bar{q}_{\beta} \, q_{\alpha})_{V-A}
      \label{operator-04}, \\
      O_{i=5,7} & = &
       (\bar{d}_{\alpha} \, b_{\alpha})_{V-A} \, \sum\limits_{q} \,
        \big( {\delta}_{i,5} + {\delta}_{i,7} \, \frac{3}{2} \, Q_{q} \big) \,
       (\bar{q}_{\beta} \, q_{\beta})_{V+A}
      \label{operator-05}, \\
      O_{i=6,8} & = &
       (\bar{d}_{\alpha} \, b_{\beta})_{V-A}  \, \sum\limits_{q}  \,
        \big( {\delta}_{i,6} + {\delta}_{i,8} \, \frac{3}{2} \, Q_{q} \big) \,
       (\bar{q}_{\beta} \, q_{\alpha})_{V+A}
      \label{operator-06},
      \end{eqnarray}
     where we use the shorthand notation $(\bar{q}_{1} \, q_{2})_{V{\pm}A}$ ${\equiv}$
     $\bar{q}_{1} \, {\gamma}_{\mu}  \, ( 1 {\pm} {\gamma}_{5} )  \, q_{2}$;
     ${\alpha}$ and ${\beta}$ are the color indices;
     $Q_{q}$ is the electric charge of quark $q$ in the unit of ${\vert}e{\vert}$;
     and $q$ ${\in}$ \{$u$, $d$, $c$, $s$, $b$\}.

     \begin{table}[t]
     \caption{Input parameters, where the central values are regarded
      as the default inputs unless specified otherwise.}
     \label{tab:input}
     \begin{ruledtabular}
     \begin{tabular}{l l l}
       \multicolumn{1}{c}{ particle mass }
     & \multicolumn{1}{c}{ decay constant }
     & \multicolumn{1}{c}{ Wolfenstein parameters } \\
        $ m_{ {\pi}^{\pm} }  $  $=$ $  139.57         $ MeV \cite{PDG.2026},
     &  $ f_{\pi}            $  $=$ $ 130.2 {\pm} 1.2 $ MeV \cite{PDG.2026},
     &  $ A                  $  $=$ $ 0.826^{+0.017}_{-0.015} $ \cite{PDG.2026}, \\
        $ m_{ {\pi}^{0} }    $  $=$ $ 134.977          $ MeV \cite{PDG.2026},
     &  $ f_{B}              $  $=$ $ 189.5 {\pm} 1.3  $ MeV \cite{PDG.2026};
     &  $ {\lambda}          $  $=$ $ 0.22517 {\pm} 0.00068 $     \cite{PDG.2026}, \\ \cline{2-2}
        $ m_{b}              $  $=$ $ 4.788  {\pm} 0.016  $ GeV \cite{PDG.2026},
     &  \multicolumn{1}{c}{life time}
     &  $ \bar{\rho}         $  $=$ $ 0.1576 ^{+0.0092}_{-0.0091} $ \cite{PDG.2026}, \\
        $ m_{ B^{\pm} }      $  $=$ $ 5279.41 {\pm} 0.07 $ MeV \cite{PDG.2026},
     &  $ {\tau}_{ B^{\pm} } $  $=$ $ 1.638 {\pm} 0.004   $ ps  \cite{PDG.2026},
     &  $ \bar{\eta}         $  $=$ $ 0.3556 ^{+0.0071}_{-0.0069} $ \cite{PDG.2026}; \\
        $ m_{ B^{0} }        $  $=$ $ 5279.63 {\pm} 0.20 $ MeV \cite{PDG.2026};
     &  $ {\tau}_{ B^{0} }   $  $=$ $ 1.5116 {\pm} 0.0031 $ ps  \cite{PDG.2026};
     &  \\ \hline
        \multicolumn{1}{c}{ parameters of wave function}
     & $ {\lambda}_{B}       $  $=$ $ 350 {\pm} 150    $ MeV \cite{NuclPhysB.606.245},
     & $ a_{2}^{\pi}         $  $=$ $ 0.25 {\pm} 0.15  $ \cite{JHEP.0605.004}.
     \end{tabular}
     \end{ruledtabular}
     \end{table}
     \begin{table}[ht]
     \caption{Numerical values of the Wilson coefficients $C_{i}$
       in the NDR scheme, with  input parameters taken as
       the $W$ boson mass $m_{W}$ $=$ $80.3625$ ${\pm}$ $0.0077$ GeV \cite{PDG.2026},
       the top quark mass $m_{t}$ $=$ $172.60$ ${\pm}$ $0.27$ GeV  \cite{PDG.2026},
       the typical scale ${\Lambda}_{h}$ $=$ $0.5$ GeV \cite{NuclPhysB.606.245},
       $ {\alpha}_{\rm em}^{-1}({\mu}_{h}) $ ${\approx}$ $ 128.8 $,
       and $ {\alpha}_{\rm em}^{-1}({\mu}_{b}) $ ${\approx}$ $ 128.6 $.
       Unless explicitly stated otherwise, the central values of
       these inputs are adopted as the default.}
     \label{tab:wcci}
     \begin{ruledtabular}
     \begin{tabular}{c rrrr}
     & \multicolumn{2}{c}{ $ {\mu}_{h} $ $=$ $ \sqrt{ m_{b}\,{\Lambda}_{h} } $ }
     & \multicolumn{2}{c}{ $ {\mu}_{b} $ $=$ $ m_{b} $ }
       \\ \cline{2-3} \cline{4-5}
     & \multicolumn{1}{c}{ LO } & \multicolumn{1}{c}{ NLO }
     & \multicolumn{1}{c}{ LO } & \multicolumn{1}{c}{ NLO } \\ \cline{2-3} \cline{4-5}
     $ C_{ 1} $ & $ ~1.220 $ & $ ~1.174 $ & $ ~1.110 $ & $ ~1.076 $ \\
     $ C_{ 2} $ & $ -0.421 $ & $ -0.348 $ & $ -0.237 $ & $ -0.173 $ \\
     $ C_{ 3} $ & $ ~0.025 $ & $ ~0.027 $ & $ ~0.012 $ & $ ~0.014 $  \\
     $ C_{ 4} $ & $ -0.058 $ & $ -0.059 $ & $ -0.032 $ & $ -0.034 $  \\
     $ C_{ 5} $ & $ ~0.011 $ & $ ~0.011 $ & $ ~0.008 $ & $ ~0.008 $  \\
     $ C_{ 6} $ & $ -0.077 $ & $ -0.080 $ & $ -0.037 $ & $ -0.039 $  \\
     $ C_{ 7} / {\alpha}_{\rm em} $ & $ -0.103 $ & $ -0.018 $ & $ -0.096 $ & $ ~0.004 $  \\
     $ C_{ 8} / {\alpha}_{\rm em} $ & $ ~0.032 $ & $ ~0.108 $ & $ ~0.014 $ & $ ~0.052 $  \\
     $ C_{ 9} / {\alpha}_{\rm em} $ & $ -0.091 $ & $ -1.432 $ & $ -0.090 $ & $ -1.297 $  \\
     $ C_{10} / {\alpha}_{\rm em} $ & $ -0.030 $ & $ ~0.455 $ & $ -0.018 $ & $ ~0.249 $
     \end{tabular}
     \end{ruledtabular}
     \end{table}

     \subsection{The $B$ ${\to}$ ${\pi}{\pi}$ decays with the naive factorization approach}
     \label{sec0202}
     The understanding of nonleptonic decays is handicapped by the
     influence of the strong interactions.
     The HME, in which the local operators $O_{i}$ are sandwiched
     between the initial $B$ meson and the final ${\pi}{\pi}$ pair,
     characterizes the transition from quarks to hadrons.
     It receives both perturbative and nonperturbative contributions,
     and thus constitutes the most challenging part of the theoretical
     calculation.

     An oversimplified phenomenological description is provided by
     the naive factorization (NF) hypothesis
     \cite{NPB.22.448,RevModPhys.47.277,NPB.100.313,NPB.133.315,
     NPB.201.315,NPB.264.371,ZPC.34.103},
     whose physical foundation rests on the ``color transparency''
     argument for energetic nonleptonic decays \cite{NPB.PS.11.325},
     together with the vacuum-insertion approximation and
     the valence quark decay assumption.
     Within the NF framework, the HMEs of four-quark operators
     are factorized into products of two HMEs of flavor-matched
     color-singlet diquark currents.
     These factorized products are subsequently parameterized
     in terms of two fundamental hadronic observables:
     the decay constant and the form factor.
     More specifically, for the $B$ ${\to}$ ${\pi}^{+}{\pi}^{-}$ decay,
     we decompose the HME of the operator $O_{1}^{u}$ as follows,
     \begin{equation}
    {\langle} \, O_{1}^{u} \, {\rangle}
     \, = \,
    {\langle} \, {\pi}^{+} \, {\pi}^{-} \, {\vert}  \, O_{1} {\vert} \, B \, {\rangle}
      \, = \,
    {\langle} \, O_{1}^{u} \, {\rangle}_{\rm f} +
    {\langle} \, O_{1}^{u} \, {\rangle}_{\rm nf}
     \label{HME-nf-o1+plus},
     \end{equation}
     where the subscripts ``f'' and ``nf'' denote the factorizable
     and nonfactorizable contributions, respectively.
     With the NF approach, the factorizable part reads
     \cite{ZPC.34.103,ZPC.29.637},
     \begin{equation}
    {\langle} \, O_{1}^{u} \, {\rangle}_{\rm f}
     \, = \,
    {\langle} \, {\pi}^{-} \, {\vert} \, (\bar{d}_{\beta}  \, u_{\beta})_{V-A} \, {\vert} \, 0 \, {\rangle} \,
    {\langle} \, {\pi}^{+} \, {\vert} \, (\bar{u}_{\alpha} \, b_{\alpha})_{V-A} \, {\vert} \, B \, {\rangle}
     \label{HME-nf-o1-f},
     \end{equation}
     \begin{equation}
    {\langle} \, {\pi}^{-} \, {\vert} \, (\bar{d}_{\beta}  \, u_{\beta})_{V-A} \, {\vert} \, 0 \, {\rangle}
     \, = \, - i \, f_{\pi}\, p_{{\pi}^{-}}^{\mu}
     \label{decay-constant},
     \end{equation}
     \begin{equation}
    {\langle} \, {\pi}^{+} \, {\vert} \, (\bar{u}_{\alpha} \, b_{\alpha})_{V-A} \, {\vert} \, B \, {\rangle}
     \, = \,
     \Big( p_{B}^{\mu} + p_{{\pi}^{+}}^{\mu}
   - \frac{ m_{B}^{2}-m_{\pi}^{2} }{q^{2}} \, q^{\mu}
     \Big) \, F_{1}^{B{\pi}} (q^{2})
   + \frac{ m_{B}^{2}-m_{\pi}^{2} }{q^{2}} \, q^{\mu} \, F_{0}^{B{\pi}} (q^{2})
     \label{form-factor},
     \end{equation}
     where $f_{\pi}$ is the pion decay constant,
     $q^{\mu}$ $=$ $p_{B}^{\mu}$ $-$ $p^{\mu}_{{\pi}^{+}}$
     is the momentum transfer,
     the $B$ ${\to}$ ${\pi}$ transition form factors
     $F_{0,1}^{B{\pi}}$ are scalar functions of $q^{2}$,
     and satisfy the condition $F_{0}^{B{\pi}}(0)$ $=$
     $F_{1}^{B{\pi}}(0)$ at $q^{2}$ $=$ $0$,
     thereby ensuring the absence of pole singularity.

     In the $1/N_{c}$ expansion, where $N_{c}$ denotes the number of colors,
     the factorizable amplitude is of order ${\cal O}( \sqrt{N_{c}} )$,
     while the nonfactorizable one is of order ${\cal O}( 1/\sqrt{N_{c}} )$ \cite{NPB.268.16}.
     Using the RG equation for the Wilson coefficients, one finds
     that the ratio \cite{Adv.Ser.Direct.HEP.15.294} is
     \begin{equation}
    {\varepsilon}_{1} \, = \,
     \frac{ {\langle} \, O_{1}^{u} \, {\rangle}_{\rm nf} }
          { {\langle} \, O_{1}^{u} \, {\rangle}_{\rm f}  } \, = \,
     {\cal O}( \frac{1}{ N_{c}^{2} } )
     \label{HME-nf-o1+e1}.
     \end{equation}

     By employing the identity relationship among the generators
     $T^{a}$ (with $a$ $=$ $1$, ${\cdots}$, $8$) of the $SU(3)$
     color group,
     \begin{equation}
     T^{a}_{i,j} \, T^{a}_{k,l} \, = \,
       \frac{1}{2} \, {\delta}_{i,l} \, {\delta}_{k,j}
     - \frac{1}{2 \, N_{c}} \, {\delta}_{i,j} \, {\delta}_{k,l}
     \label{eq:su3-color},
     \end{equation}
     where the subscripts $i$, $j$, $k$, $l$ are color indices,
     the operator $O_{2}^{u}$ can be equivalently recast as follows:
     \begin{eqnarray}
     O_{2}^{u} & = &
    (\bar{u}_{\alpha} \, b_{\beta}  )_{V-A} \,
    (\bar{d}_{\beta}  \, u_{\alpha} )_{V-A}
     \nonumber \\ & = &
     \frac{1}{N_{c}} \,
    (\bar{u}_{\alpha} \, b_{\alpha}  )_{V-A} \,
    (\bar{d}_{\beta}  \, u_{\beta} )_{V-A}
     + 2 \,
    (\bar{u} \, T^{a} \, b )_{V-A} \,
    (\bar{d} \, T^{a} \, u )_{V-A}
     \nonumber \\ & = &
    \frac{1}{N_{c}} \, O_{1}^{u}
     + 2 \, \widetilde{O}_{2}^{u}
     \label{fierz-O2},
     \end{eqnarray}
     The operator $\widetilde{O}_{2}^{u}$ is composed of
     two color-octet currents:
     \begin{equation}
     \widetilde{O}_{2}^{u} \, = \,
    (\bar{u} \, T^{a} \, b )_{V-A} \,
    (\bar{d} \, T^{a} \, u )_{V-A}
     \label{operator-o2-color-octet}.
     \end{equation}
     The evaluations in Ref. \cite{Adv.Ser.Direct.HEP.15.294}
     show that the ratio is
     \begin{equation}
    {\varepsilon}_{8} \, = \, 2 \,
     \frac{ {\langle} \, \widetilde{O}_{2}^{u} \, {\rangle} }
          { {\langle} \, O_{1}^{u} \, {\rangle}_{\rm f}  } \, = \,
     {\cal O}( \frac{1}{ N_{c} } )
     \label{HME-nf-o2+e8}.
     \end{equation}
     The contribution of ${\langle} \, \widetilde{O}_{2}^{u} \, {\rangle}$,
     whose color configuration does not match that of the
     participating hadrons under the vacuum-insertion approximation,
     is completely discarded by the NF approach.

     There exists, in principle, another type of factorizable contribution,
     the so-called ``weak annihilation (WA) contribution''.
     For example, by applying the Fierz rearrangement to the
     products of two color-mismatched diquark currents,
     one obtains the following factorized form:
     \begin{eqnarray}
    {\langle} \, O_{2}^{u} \, {\rangle}_{a}
     & = &
    {\langle} \, {\pi}^{+} \, {\pi}^{-} \, {\vert} \,
    ( \bar{u}_{\alpha} \, b_{\beta} )_{V-A} \,
    ( \bar{d}_{\beta} \, u_{\alpha} )_{V-A} \,
    {\vert} \, B \, {\rangle}_{a}
     \nonumber \\ & = &
    {\langle} \, {\pi}^{+} \, {\pi}^{-} \, {\vert} \,
    ( \bar{u}_{\alpha} \, u_{\alpha} )_{V-A} \,
    ( \bar{d}_{\beta} \, b_{\beta} )_{V-A} \,
    {\vert} \, B \, {\rangle}_{a}
     \nonumber \\ & = &
    {\langle} \, {\pi}^{+} \, {\pi}^{-} \, {\vert} \,
    ( \bar{u}_{\alpha} \, u_{\alpha} )_{V-A} \,
    {\vert} \, 0 \, {\rangle}\,
    {\langle} \, 0 \, {\vert} \,
    ( \bar{d}_{\beta} \, b_{\beta} )_{V-A} \,
    {\vert} \, B \, {\rangle}
     \label{HME-nf-o2-wa},
     \end{eqnarray}
     where the subscript ``$a$'' denotes the WA contribution.
     The WA contributions are formally disregarded in the
     NF approach on the basis of helicity suppression
     arguments \cite{PLB.90.455}
     or the $1/m_{b}$ power counting rules in the heavy quark
     limit \cite{NuclPhysB.591.313}.
     In addition, as stated in Ref.~\cite{Adv.Ser.Direct.HEP.15.294},
     the ``color transparency'' argument precludes the possibility
     of significant final state interactions for the energetic
     $B$ ${\to}$ ${\pi}{\pi}$ decays.

     To summarize, within the NF scheme, no attempt is made to include
     nonfactorizable contributions, terms beyond the $1/N_{c}$ expansion,
     and WA amplitudes.
     Adopting the conventions for the HMEs defined in
     Eqs.(\ref{HME-nf-o1-f}), (\ref{decay-constant}), and
     (\ref{form-factor}), the decay amplitudes for the
     $B$ ${\to}$ ${\pi}{\pi}$ processes
     can be expressed as follows:
     \begin{equation}
    {\cal F}_{{\pi}{\pi}} \, = \,
     \frac{G_{F}}{\sqrt{2}} \,
    {\langle} \, {\pi}^{+} \, {\pi}^{-} \, {\vert}  \, O_{1}^{u} \, {\vert} \, B \, {\rangle}_{f}
     \, = \,
     - i \, \frac{G_{F}}{\sqrt{2}} \, (m_{B}^{2}-m_{\pi}^{2}) \, f_{\pi} \,
      F_{0}^{B{\pi}} ( m_{\pi}^{2})
     \label{eq:amp-pi-pi},
     \end{equation}
     \begin{equation}
    {\cal M}^{f}_{+-} \, = \,
    {\cal F}_{{\pi}{\pi}}  \,
     \big\{ V_{ub}\,V_{ud}^{\ast} \, a_{1}
          - V_{tb}\,V_{td}^{\ast} \, \big(
            a_{4}+a_{10}+a_{6}\, r_{\pi} + a_{8} \, r_{\pi} \big) \big\}
     \label{eq:amp-pim-pip},
     \end{equation}
     \begin{equation}
    {\cal M}^{f}_{-0} \, = \,
     \frac{ {\cal F}_{{\pi}{\pi}} }{\sqrt{2}} \,
     \big\{ V_{ub}\,V_{ud}^{\ast} \, \big( a_{1} + a_{2} \big)
         -  V_{tb}\,V_{td}^{\ast} \, \frac{3}{2}\, \big(
       a_{9} + a_{10} - a_{7} + a_{8} \, r_{\pi} \big) \big\}
     \label{eq:amp-pim-piz},
     \end{equation}
     \begin{equation}
    {\cal M}^{f}_{00} \, = \,
   -{\cal F}_{{\pi}{\pi}}  \,
     \big\{  V_{ub}\,V_{ud}^{\ast} \, a_{2}
           + V_{tb}\,V_{td}^{\ast} \, \big(
             a_{4} + a_{6} \, r_{\pi} + \frac{3}{2}\, a_{7}
           - \frac{1}{2} \, a_{8} \, r_{\pi}
           - \frac{3}{2} \, a_{9}
           - \frac{1}{2} \, a_{10} \big) \big\}
     \label{eq:amp-piz-piz},
     \end{equation}
     where the superscript ``$f$'' attached to ${\cal M}^{f}_{ij}$
     designates the factorizable amplitude, and the subscripts ``$i$''
     and ``$j$'' specify the isospin $z$-component of the pions.
     The coefficient is given by the following combination of
     Wilson coefficients:
     \begin{equation}
     a_{i} \, = \, C_{i} + \frac{1}{N_{c}} \, C_{j}
     \label{eq:ai-nf},
     \end{equation}
     with $j$ $=$ $i$ $-$ $1$ for even $i$, and
     $j$ $=$ $i$ $+$ $1$ for odd $i$.
     The chiral factor $r_{\pi}$ $=$
     $m_{\pi}^{2}/[ \overline{m}_{b}({\mu}) \, \overline{m}_{q}({\mu}) ]$
     is computed via the RG equation for the ``running mass'',
     with inputs
     $\overline{m}_{b}(\overline{m}_{b})$ $=$ $4.1859$ ${\pm}$ $0.0034$ GeV and
     $\overline{m}_{q}(\text{2\,GeV})$ $=$ $\frac{1}{2}\,(\overline{m}_{u}+\overline{m}_{d})$
     $=$ $3.49$ ${\pm}$ $0.04$ MeV \cite{PDG.2026}.

     The NF scheme suffers from several inherent drawbacks.
     The first concerns the renormalization scale dependence.
     The HME in Eq.(\ref{eq:amp-pi-pi}) is independent of the
     renormalization scale ${\mu}$,
     which implies that the ${\mu}$ dependence of the Wilson coefficients
     remains uncompensated. Consequently, the resulting decay
     amplitudes are scale ambiguous and hence unphysical.
     The second issue is the complete neglect of nonfactorizable
     contributions. This omission is likely responsible for the
     inadequacy of the NF approach, particularly in describing
     color-suppressed $B$ decays and $CP$ asymmetries;
     consequently, the coefficient $a_{2}$
     becomes highly sensitive to any additional nonfactorizable
     effects. The third deficiency is that
     the decay amplitudes are entirely devoid of strong phases,
     even though such phases are crucial for reproducing the
     observed $CP$ asymmetries in exclusive nonleptonic $B$ decays.

     \subsection{The $B$ ${\to}$ ${\pi}{\pi}$ decays with the QCDF approach}
     \label{sec0203}

     The aforementioned shortcomings of the NF scheme are largely
     addressed by the QCDF approach \cite{NuclPhysB.591.313,
     NuclPhysB.606.245,PhysLettB.488.46,PhysLettB.509.263,
     PhysRevD.64.014036}.
     Following the factorization approach for exclusive hadronic
     processes proposed in Ref. \cite{PRD.22.2157} and
     the master QCDF formula \cite{NuclPhysB.591.313},
     HMEs can be approximated,
     up to the power corrections of order $1/m_{b}$,
     as convolutions of scattering kernels with
     wave functions of the participating mesons:
     \begin{eqnarray}
    {\langle} \, {\pi}_{1} \, {\pi}_{2} \, {\vert} \, O_{i} \, {\vert} \, B \, {\rangle}
     & = &
     F_{0}^{B{\pi}} ( m_{\pi}^{2} ) \, f_{\pi} \,
     \Big\{ {\int} dy \, {\cal T}^{I}(y) \, {\Phi}_{{\pi}_{1}}(y)
          + {\int} dz \, {\cal T}^{I}(z) \, {\Phi}_{{\pi}_{2}}(z) \Big\}
     \nonumber \\ & + &
     f_{B} \, f_{\pi}^{2} \,
    {\int} dx \, dy \, dz \, {\cal T}^{II}(x,y,z) \,
    {\Phi}_{B}(x) \, {\Phi}_{{\pi}_{1}}(y) \, {\Phi}_{{\pi}_{2}}(z)
     \label{QCDF-master-formula},
     \end{eqnarray}
     where the nonperturbative quantities $F_{0}^{B{\pi}}$ and
     $f_{\pi}$ are the form factor and pion decay constant specified
     in Eq.(\ref{form-factor}) and Eq.(\ref{decay-constant}),
     respectively.
     The functions ${\cal T}^{I}$ and ${\cal T}^{II}$ are the
     scattering kernels that account for the hard gluon
     exchange interactions among the quarks and are reliably
     calculable in perturbative QCD.
     ${\Phi}_{i}$ is the universal meson wave function.
     The variables $x$, $y$, and $z$ denote the longitudinal
     momentum fractions of valence quarks.

     The calculation yields the effective QCDF coefficients $a_{i}^{p}$
     \cite{NuclPhysB.606.245}:
     \begin{equation}
     a_{i}^{p} \, = \, a_{i,I}^{p} + a_{i,II},
     \label{HME-O1-factorizable}
     \end{equation}
     where $a_{i,I}^{p}$ (with $p$ $=$ $u$, $c$)
     correspond to the form factor terms
     in Eq.(\ref{QCDF-master-formula}) and contain the
     NF contributions as well as the vertex and penguin
     corrections, while $a_{i,II}$ arises from the
     hard spectator scattering (HSS) interactions.
     The explicit analytic expressions for the QCDF coefficients
     $a_{i}^{p}$ at NLO are available in Ref. \cite{NuclPhysB.606.245},
     and their numerical values for $a_{i,I}^{p}$
     are listed in Table~\ref{value-aiI}.
     It is evident that the QCDF coefficients $a_{i,I}^{p}$
     (or $a_{i}^{p}$) at NLO contain nonzero imaginary parts.
     The partial strong rescattering phases are successfully
     reproduced by taking into account the nonfactorizable
     contributions within the QCDF approach.
     It has been shown in Refs. \cite{NuclPhysB.591.313,
     PhysLettB.488.46,PhysLettB.509.263,PhysRevD.64.014036}
     that the QCDF coefficients $a_{i,I}^{p}$ are
     gauge invariant, infrared finite, and independent of the
     renormalization scale at order ${\alpha}_{s}$.
     The residual scale dependence observed in the numerical results
     of Table~3 in Ref. \cite{NuclPhysB.606.245} is entirely
     attributable to the truncation of the perturbative series.
     In this work, we adopt the NLO coefficients given in
     Ref. \cite{NuclPhysB.606.245} for the subsequent calculations.
     The detailed expressions and values for the QCDF coefficients
     $a_{i}^{p}$ at NNLO for the $B$ ${\to}$ ${\pi}{\pi}$ decays can
     be found in Refs. \cite{NuclPhysB.832.109,JHEP.2020.04.055}.

     \begin{table}[t]
     \caption{Numerical values of the NLO coefficients $a_{i,I}^{p}$
       at the scale ${\mu}$ $=$ $m_{b}$.}
     \label{value-aiI}
     \begin{ruledtabular}
     \begin{tabular}{rrr}
       $ a_{1,I}^{p} $ $=$ $ 1.052 $ $+$ $ i \, 0.017 $
     & $ a_{2,I}^{p} $ $=$ $ 0.027 $ $-$ $ i \, 0.080 $
     & $ a_{3,I}^{p} $ $=$ $ 0.007 $ $+$ $ i \, 0.002 $ \\
       $ -a_{4,I}^{u} $ $=$ $ 0.036 $ $+$ $ i \, 0.015 $
     & $ -a_{4,I}^{c} $ $=$ $ 0.039 $ $+$ $ i \, 0.004 $
     & $ -a_{5,I}^{p} $     $=$ $ 0.006 $ $+$ $ i \, 0.003 $ \\
       $ -a_{6,I}^{u} $ $=$ $ 0.042 $ $+$ $ i \, 0.014 $
     & $ -a_{6,I}^{c} $ $=$ $ 0.045 $ $+$ $ i \, 0.003 $
     & $ a_{7,I}^{p}/{\alpha}_{\rm em} $ $=$ $ 0.022 $ $+$ $ i \, 0.001 $ \\
       $ a_{8,I}^{u}/{\alpha}_{\rm em} $ $=$ $ 0.053 $ $-$ $ i \, 0.010 $
     & $ a_{8,I}^{c}/{\alpha}_{\rm em} $ $=$ $ 0.051 $ $-$ $ i \, 0.004 $
     & $ -a_{9,I}^{p} /{\alpha}_{\rm em} $ $=$ $ 1.212 $ $-$ $ i \, 0.001 $ \\
       $ -a_{10,I}^{u}/{\alpha}_{\rm em} $ $=$ $ 0.176 $ $+$ $ i \, 0.003 $
     & $ -a_{10,I}^{c}/{\alpha}_{\rm em} $ $=$ $ 0.178 $ $-$ $ i \, 0.003 $
     \end{tabular}
     \end{ruledtabular}
     \end{table}

     As demonstrated in Refs. \cite{PhysRevD.65.074001,
     NuclPhysB.675.333,NuclPhysB.832.109,PhysRevD.80.114008,
     PhysRevD.90.054019,PhysRevD.74.014003,PhysRevD.90.014029,
     PhysRevD.91.114019,PhysRevD.103.056006,ChinPhysC.46.123103,
     PhysRevD.107.013004,PhysRevD.110.056025}, the WA contributions
     are numerically significant and indispensable for both the
     branching ratios and the $CP$ asymmetries in the $B$ ${\to}$
     ${\pi}{\pi}$ decays, irrespective of whether the QCDF or
     PQCD approach is employed.
     In this work, we adopt the prescription of Ref. \cite{NuclPhysB.606.245}
     for the WA HMEs in terms of convolutions of scattering kernels
     with light-cone distribution amplitudes (LCDAs) of the participating
     mesons, and take the WA contributions into account accordingly.
     The explicit expressions for the WA amplitudes
     ${\cal M}^{a}_{{\pi}{\pi}}$ can be found in
     Refs. \cite{PhysRevD.65.074001,NuclPhysB.675.333}
     and are not reproduced here for brevity.
     For consistency throughout this work, we have revised the
     parametric expressions for the basic WA building blocks
     originally presented in Ref. \cite{NuclPhysB.606.245}
     by including the truncation at $n$ $=$ $2$ in the
     Gegenbauer polynomial expansion of the pion twist-2 LCDAs,
     in analogy with the treatment of the vertex corrections and
     HSS contributions to the QCDF coefficients $a_{i}^{p}$.
     The modified expressions for the WA building blocks are
     as follows:
     \begin{eqnarray}
     A^{i}_{1,2} & = & {\pi} \, {\alpha}_{s} \, \{
       2 \, r_{\pi}^{2} \, X_{A}^{2}
     + 6 \, [ 3 \, X_{A} + {\pi}^{2} - 12
     \nonumber \\ & & \quad
     + \, ( 21 \, X_{A} + 42 \, {\pi}^{2} - 464 ) \, a_{2}^{\pi}
     \nonumber \\ & & \quad
     + \, ( 18 \, X_{A}+ 2916 \, {\pi}^{2} - 28 \, 827  ) \, (a_{2}^{\pi})^{2} ] \}
     \label{WA-ai-12},
     \end{eqnarray}
     \begin{equation}
     A^{f}_{3} \, = \, 12 \, {\pi} \, {\alpha}_{s} \, r_{\pi} \, X_{A} \,
      \{ 2 \, X_{A} - 1 + ( 12 \, X_{A} - 31 ) \, a_{2}^{\pi} \}
     \label{WA-af-3},
     \end{equation}
     where the quantities $A^{i,f}_{k}$ have the same
     meanings as those in Eq.(62) of Ref. \cite{NuclPhysB.606.245}.
     The Gegenbauer moment $a_{2}^{\pi}$ is closely related to
     the pion twist-2 LCDAs, which are conventionally expanded in
     terms of the Gegenbauer polynomials as \cite{JHEP.0605.004}:
     \begin{equation}
    {\Phi}_{\pi}(x) \, = \, 6 \, x \, \bar{x} \,
     \sum\limits_{n=0} a_{n}^{\pi} \, C_{n}^{3/2}(x-\bar{x})
     \label{eq:pion-distribution amplitudes},
     \end{equation}
     where $x$ and $\bar{x}$ $=$ $1$ $-$ $x$ denote the
     longitudinal momentum fractions carried by the valence
     quarks inside the pion, with $a_{0}^{\pi}$ $=$ $1$
     and $a_{1}^{\pi}$ $=$ $0$ by virtue of isospin symmetry
     and $G$-parity.
     The quantity $X_{A}$ is defined as follows \cite{NuclPhysB.606.245,NuclPhysB.675.333}:
     \begin{equation}
     X_{A}
     \, = \, {\int}_{0}^{1} \frac{dx}{x}
     \, = \, ( 1 + {\rho}_{A} \, e^{ i \, {\phi}_{A} } ) \,
          {\ln}\frac{ m_{B} }{ {\Lambda}_{h} }
     \label{WA-Xa},
     \end{equation}
     where the scale is set to ${\Lambda}_{h}$ $=$ $0.5$ GeV,
     and ${\phi}_{A}$ is an arbitrary phase that may arise from
     soft rescattering effects.
     As proposed in Refs.~\cite{NuclPhysB.606.245,NuclPhysB.675.333},
     the condition ${\rho}_{A}$ $<$ $1$ is generally imposed to
     guarantee the reliability of the convolution treatment for
     the WA HMEs within the QCDF framework.

     Besides the WA endpoint divergence discussed above, an
     analogous endpoint divergence also appears in the HSS
     amplitudes.
     This divergence is handled using a similar parametrization
     scheme \cite{NuclPhysB.606.245,NuclPhysB.675.333}:
     \begin{equation}
     X_{H} \, = \, ( 1 + {\rho}_{H} \, e^{ i \, {\phi}_{H} } ) \,
          {\ln}\frac{ m_{B} }{ {\Lambda}_{h} }
     \label{HS-Xh},
     \end{equation}
     with the same phenomenological interpretation for the parameters
     ${\rho}_{H}$ and ${\phi}_{H}$ as in the WA case.
     It should be noted, however, that the parameters $X_{H}$ and
     $X_{A}$ are in principle independent.
     They are associated with convolution integrals over pion
     LCDAs of different twists and correspond to distinct
     topological structures, and therefore should not be
     identified with each other.
     Moreover, the parameters $X_{A,H}$ can be further split
     into multiple parameters according to different
     topological structures (see, {\em e.g.}, Refs.
     \cite{PhysRevD.90.054019,PLB.740.56,PLB.743.444,
     PhysRevD.91.074026,JPG.43.105004}) or,
     alternatively, according to different final states
     (see, {\em e.g.}, Refs. \cite{NuclPhysB.675.333,
     PhysRevD.80.114008,PhysRevD.88.014043}).
     Such arbitrariness inevitably introduces considerable
     theoretical uncertainties and significantly undermines
     the predictive power of the QCDF approach.
     Since only the ${\pi}{\pi}$ final states are considered here,
     and in view of the constraints from isospin symmetry
     as well as the limited number of available experimental data,
     we will adopt the convention of Ref.~\cite{NuclPhysB.606.245}
     to reduce the number of free parameters.
     Thus, for the time being, we take $X_{H}$ $=$ $X_{A}$.

     \subsection{The soft COME contributions to the $B$ ${\to}$ ${\pi}{\pi}$ decays}
     \label{sec0204}

     From the detailed analysis of the nonfactorizable contributions
     to the $B$ ${\to}$ $D{\pi}$ process in Ref.~\cite{NuclPhysB.591.313},
     it is evident that the hard COME contributions of the operator
     $\widetilde{O}_{2}^{c}$ (denoted as $O_{8}$ in Ref.~\cite{NuclPhysB.591.313})
     correspond to the vertex corrections encoded in the
     coefficients $a_{i,I}^{c}$.
     For the $B$ ${\to}$ ${\pi}{\pi}$ decays, the hard COME
     contributions from the $\widetilde{O}_{2}^{p}$-like operators
     have been consistently incorporated into the effective QCDF
     coefficient $a_{i}^{p}$ in Eq.(\ref{HME-O1-factorizable}).
     However, the soft COME contributions have not yet been
     thoroughly investigated.
     Although it was estimated in
     Ref.~\cite{NuclPhysB.591.313} that they are power
     suppressed relative to the hard ones in the HSS amplitudes,
     their numerical importance cannot be excluded {\it a priori} for
     realistic $B$ meson decays, given that $m_{b}$ is not
     asymptotically large.
     It is therefore worthwhile to investigate whether such soft
     nonperturbative effects could play a role in the
     $B$ ${\to}$ ${\pi}{\pi}$ decays.
     Given the nonperturbative nature of soft COME contributions,
     we do not attempt to evaluate them
     using nonperturbative methods, such as the QCD sum rules
     employed in Ref.~\cite{NPB.389.534}.
     Instead, following the strategy proposed in Refs.
     \cite{PhysRevD.107.013004,Adv.Ser.Direct.HEP.15.294},
     we parameterize the ratio of the soft COME to the
     factorizable color-singlet HME in terms of a complex
     parameter $X_{c}$, defined as
     \begin{equation}
     X_{c} \, = \,
     \frac{ {\langle} \, \widetilde{O}_{2}^{u} \, {\rangle}_{\rm soft} }
          { {\langle} \, O_{1}^{u} \, {\rangle}_{\rm f}  }
     \, = \, {\rho}_{c} \, e^{ i \, {\phi}_{c} }
     \label{O2-XC},
     \end{equation}
     where ${\rho}_{c}$ and ${\phi}_{c}$ are assumed to be universal
     for the $B$ ${\to}$ ${\pi}{\pi}$ decays, based on isospin
     symmetry in the strong interactions.
     Consequently, the decay amplitudes arising solely from
     soft COME contributions can be expressed as:
     \begin{equation}
    {\cal M}^{c}_{{\pi}{\pi}} \, = \, 2 \,  X_{c} \,
    {\cal M}^{f}_{{\pi}{\pi}}( a_{i} \, {\to} \, C_{j}^{\rm NLO}({\mu}_{f}) )
     \label{eq:amp-octet},
     \end{equation}
     where $j$ $=$ $i$ $-$ $1$ for even $i$, and
     $j$ $=$ $i$ $+$ $1$ for odd $i$; and
     the factorizable decay amplitudes ${\cal M}^{f}_{{\pi}{\pi}}$
     have been given in Eqs.(\ref{eq:amp-pim-pip}),
     (\ref{eq:amp-pim-piz}), and (\ref{eq:amp-piz-piz}).
     The factorization scale ${\mu}_{f}$ serves to separate
     perturbative and nonperturbative contributions.
     In analogy with the treatment of the scale in WA amplitudes
     within the QCDF framework, we adopt the approximation
     ${\mu}_{f}$ $=$ ${\mu}_{h}$ in this work to bridge the
     hard and soft contributions, and we do not dwell on the
     ${\mu}_{f}$ scale dependence for the moment.
     As can be seen from Eq.(\ref{eq:amp-octet}), the soft COME
     amplitude for the $B$ ${\to}$ ${\pi}^{0}{\pi}^{0}$
     (${\pi}^{+}{\pi}^{-}$) decay is proportional to the
     coefficient $C_{1}$ ($C_{2}$), and thus would significantly
     (mildly) affect the branching ratio and $CP$ asymmetries.
     This observation suggests a possible alternative mechanism
     for accommodating the ``${\pi}{\pi}$ puzzle'' within
     the QCDF approach.

     Based on the above analysis, the decay amplitudes, including
     the hard and soft nonfactorizable COME effects as well as
     the WA contributions, can be written as
     \begin{equation}
       {\cal M}_{{\pi}{\pi}} \, = \,
       {\cal M}^{f}_{{\pi}{\pi}}
     + {\cal M}^{a}_{{\pi}{\pi}}
     + {\cal M}^{c}_{{\pi}{\pi}}
     \label{eq:amp-singlet+octet+wa}.
     \end{equation}

     \section{Numerical results and discussion}
     \label{sec03}
     To quantitatively assess the contributions from the factorizable,
     nonfactorizable, WA, and soft COME amplitudes, we employ the
     least-squares method to construct estimators for the
     unknown parameters by making full use of the PDG data.
     The chi-square function is defined as
     \begin{equation}
    {\chi}_{i} \, = \,
     \frac{  y_{i,{\rm theo.}} - y_{i,{\rm exp.}} }{ {\sigma}_{i} }
     \label{eq:chix},
     \end{equation}
     \begin{equation}
    {\chi}^{2} \, = \,
     \sum\limits_{i} {\chi}_{i}^{2}
     \label{eq:chi2},
     \end{equation}
     where $y_{i}$ denotes the branching ratios and $CP$
     asymmetries, and ${\sigma}_{i}$ is
     the corresponding experimental uncertainty.
     The experimental data $y_{i,{\rm exp.}}$ and errors
     ${\sigma}_{i}$ are listed in Table~\ref{tab:exp-data}.
     The closer the absolute value of ${\chi}_{i}$ is to zero,
     the better the agreement between the theoretical
     predictions and the experimental results.

     The branching ratio is computed as:
     \begin{equation}
    {\cal B} \, = \,
     \frac{ p_{\rm cm} }{ 16\, {\pi} } \,
     \frac{ {\tau}_{B} }{ m_{B}^{2} } \,
     \big( {\vert} {\cal M}_{{\pi}{\pi}} {\vert}^{2}
     +     {\vert} \overline{\cal M}_{{\pi}{\pi}} {\vert}^{2}
     \big)
     \label{eq:br},
     \end{equation}
     where $p_{\rm cm}$ is the pion momentum in the rest frame
     of the decaying $B$ meson,
     ${\tau}_{B}$ is the $B$ meson lifetime, and
     $\overline{\cal M}_{{\pi}{\pi}}$ denotes the Hermitian
     conjugate of the amplitude given in
     Eq.(\ref{eq:amp-singlet+octet+wa}).

     For charged $B$ decays, the direct $CP$ asymmetry is given by
     \begin{equation}
    {\cal A}_{CP} \, = \,
     \frac{ {\vert} {\cal M}( B^{-} \, {\to} \, {\pi}^{-} \, {\pi}^{0} ) {\vert}^{2} -
            {\vert} {\cal M}( B^{+} \, {\to} \, {\pi}^{+} \, {\pi}^{0} ) {\vert}^{2} }
          { {\vert} {\cal M}( B^{-} \, {\to} \, {\pi}^{-} \, {\pi}^{0} ) {\vert}^{2} +
            {\vert} {\cal M}( B^{+} \, {\to} \, {\pi}^{+} \, {\pi}^{0} ) {\vert}^{2} }
     \label{acp-asymmetry}.
     \end{equation}

     For neutral $B$ decays, the time-dependent $CP$ asymmetry is given by
     \begin{eqnarray}
    {\cal A}_{CP} & = &
     \frac{ {\vert} {\cal M}( \overline{B}^{0} \, {\to} \, {\pi} \, {\pi} ) {\vert}^{2} -
            {\vert} {\cal M}( B^{0} \, {\to} \, {\pi} \, {\pi} ) {\vert}^{2} }
          { {\vert} {\cal M}( \overline{B}^{0} \, {\to} \, {\pi} \, {\pi} ) {\vert}^{2} +
            {\vert} {\cal M}( B^{0} \, {\to} \, {\pi} \, {\pi} ) {\vert}^{2} }
     \nonumber \\ & = &
    {\cal S}_{CP} \, {\sin}({\Delta}m_{d} \, t)
   -{\cal C}_{CP} \, {\cos}({\Delta}m_{d} \, t)
     \label{time-dependent-acp-asymmetry},
     \end{eqnarray}
     where ${\Delta}m_{d}$ is the mass difference between
     the two physical $B_{d}$ meson eigenstates, corresponding
     to the $B^{0}$-$\overline{B}^{0}$ oscillation frequency.
     The mixing-induced and direct $CP$ asymmetry parameters,
     ${\cal S}_{CP}$ and ${\cal C}_{CP}$, are respectively
     defined as
     \begin{equation}
    {\lambda}_{CP} \, = \,
     \frac{ V_{td} \, V_{tb}^{\ast} }{ V_{td}^{\ast} \, V_{tb} } \,
     \frac{ {\cal M}( \overline{B}^{0} \, {\to} \, {\pi} \, {\pi} ) }
          { {\cal M}( B^{0} \, {\to} \, {\pi} \, {\pi} ) }
     \label{cp-asymmetry-lambda},
     \end{equation}
     \begin{equation}
    {\cal S}_{CP} \, = \,
     \frac{ 2 \, {\rm Im} ( {\lambda}_{CP} ) }
          { 1 + {\vert} {\lambda}_{CP} {\vert}^{2} }
     \label{scp-asymmetry},
     \end{equation}
     \begin{equation}
    {\cal C}_{CP} \, = \,
     \frac{ 1 - {\vert} {\lambda}_{CP} {\vert}^{2} }
          { 1 + {\vert} {\lambda}_{CP} {\vert}^{2} }
     \label{ccp-asymmetry}.
     \end{equation}

     The free nonperturbative parameters in our fit include
     the form factor $F_{0}^{B{\pi}}$,
     the WA parameter $X_{A}$ $=$ $X_{H}$
     (parametrized by ${\rho}_{A}$ and ${\phi}_{A}$),
     and the soft COME parameter $X_{c}$
     (parametrized by ${\rho}_{c}$ and ${\phi}_{c}$).
     All other parameters are fixed to their central values as listed
     in Table~\ref{tab:input}.
     This choice is based on the following considerations.
     (1)
     On the one hand, the available experimental measurements
     are limited, both in terms of the number of observables and
     their precision.
     For example, the relative uncertainties given by the PDG group
     are $13\%$ for the branching ratio ${\cal B}( {\pi}^{0} {\pi}^{0} )$,
     and $78\%$ and $400\%$ for the $CP$ asymmetries
     ${\cal C}_{CP}( {\pi}^{0} {\pi}^{0} )$ and
     ${\cal A}_{CP}( {\pi}^{-} {\pi}^{0} )$,
     respectively (see Table~\ref{tab:exp-data}).
     Given these circumstances, it is challenging to impose
     tight constraints on the theoretical model parameters, and
     an excessive number of free parameters would inevitably
     introduce additional uncertainties into the fit.
     (2)
     On the other hand, except for the Gegenbauer moment
     $a_{2}^{\pi}$ and the inverse moment of the $B$
     meson LCDA ${\lambda}_{B}$, all other input parameters
     in Table~\ref{tab:input} have already been determined
     with high precision, and the existing $B$ ${\to}$ ${\pi}{\pi}$
     decay measurements alone are insufficient to further improve
     their constraints.
     To explore the impact of different assumptions,
     we consider the following three scenarios.
     \begin{enumerate}
     \item[(1)]
     Scenario S1 (``universal WA''):
     In this scenario, we set the soft COME amplitude
     ${\cal M}^{c}_{{\pi}{\pi}}$ $=$ $0$ in Eq.(\ref{eq:amp-singlet+octet+wa})
     and $X_{H}$ $=$ $X_{A}$,
     which corresponds to the conventional QCDF prescription.
     The free parameters are $F_{0}^{B{\pi}}$,
     ${\rho}_{A}$, and ${\phi}_{A}$.
     \item[(2)]
     Scenario S2 (``universal soft COME''):
     In this scenario, we set the WA amplitude
     ${\cal M}^{a}_{{\pi}{\pi}}$ $=$ $0$
     in Eq.(\ref{eq:amp-singlet+octet+wa})
     and the QCDF coefficient $a_{i,II}$ $=$ $0$
     in Eq.(\ref{HME-O1-factorizable}), thus excluding
     the HSS contributions.
     The free parameters are $F_{0}^{B{\pi}}$,
     ${\rho}_{c}$, and ${\phi}_{c}$.
     \item[(3)]
     Scenario S3 (``combined''):
     In this scenario, we take into account both the WA and soft
     COME amplitudes, as well as the interference between them.
     The free parameters are $F_{0}^{B{\pi}}$,
     ${\rho}_{A}$, ${\phi}_{A}$, ${\rho}_{c}$, and ${\phi}_{c}$.
     This is our favored scenario.
     \end{enumerate}

     Scenarios S1 and S2 serve as control cases for comparison
     with our favored Scenario S3.
     The goodness-of-fit for these two scenarios is substantially
     worse than that for Scenario S3.
     Given such poor fits, the best-fit parameter values obtained
     in Scenarios S1 and S2 are essentially meaningless, and
     no serious physical conclusions can be drawn from Scenarios
     S1 and S2, other than that they fail to account for the
     experimental data.
     Accordingly, for these two scenarios we report only the
     corresponding qualitative conclusions, and we refrain from
     extensive analysis and discussion of their fitted results,
     since a detailed discussion of poorly fitted scenarios
     would be of limited value.

     \subsection{Scenario S1}
     \label{sec0301}
\begin{figure}[h]
     \includegraphics[width=0.4\textwidth]{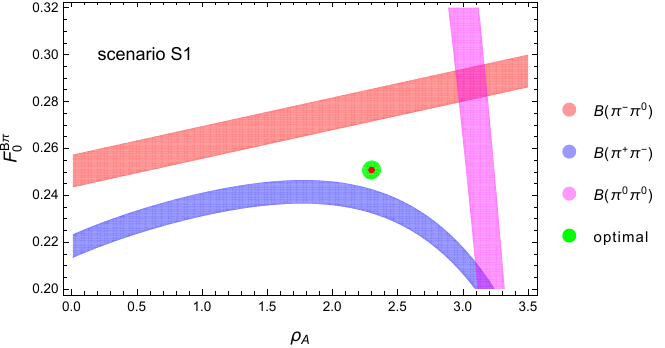} \qquad \qquad
     \includegraphics[width=0.4\textwidth]{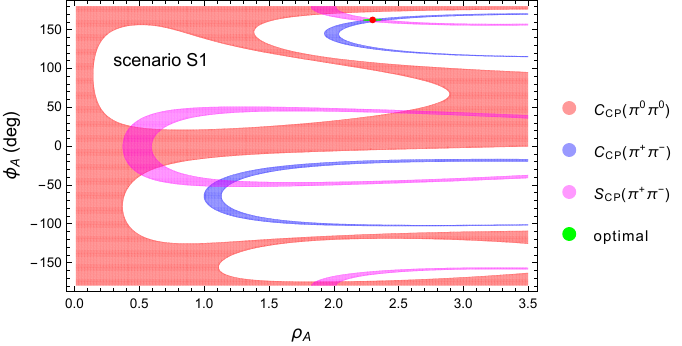}
     \caption{Constraints on the S1 parameters.
       The bands represent the parameter regions allowed by
       the PDG data within the quoted uncertainties,
       the green oval corresponds to the $1\,{\sigma}$ range of the
       fitted parameters, and the red dot indicates the central
       value of the best-fit solution.}
     \label{fig:s1constraints}
     \end{figure}

     The constraints on the form factor $F_{0}^{B{\pi}}$
     and the parameter ${\rho}_{A}$ are mainly derived from the
     measurements of branching ratios.
     Since $CP$ asymmetries are defined as ratios of the squared
     amplitudes, they exhibit a very weak dependence on the form
     factors. Fig.~\ref{fig:s1constraints} displays the respective
     constraints imposed by these two types of observables on
     the parameter space. It is clearly seen that
     the resulting fits cannot yield a consistent and physically
     satisfactory interpretation of the measured branching
     ratios.

\begin{figure}[h]
     \includegraphics[width=0.5\textwidth]{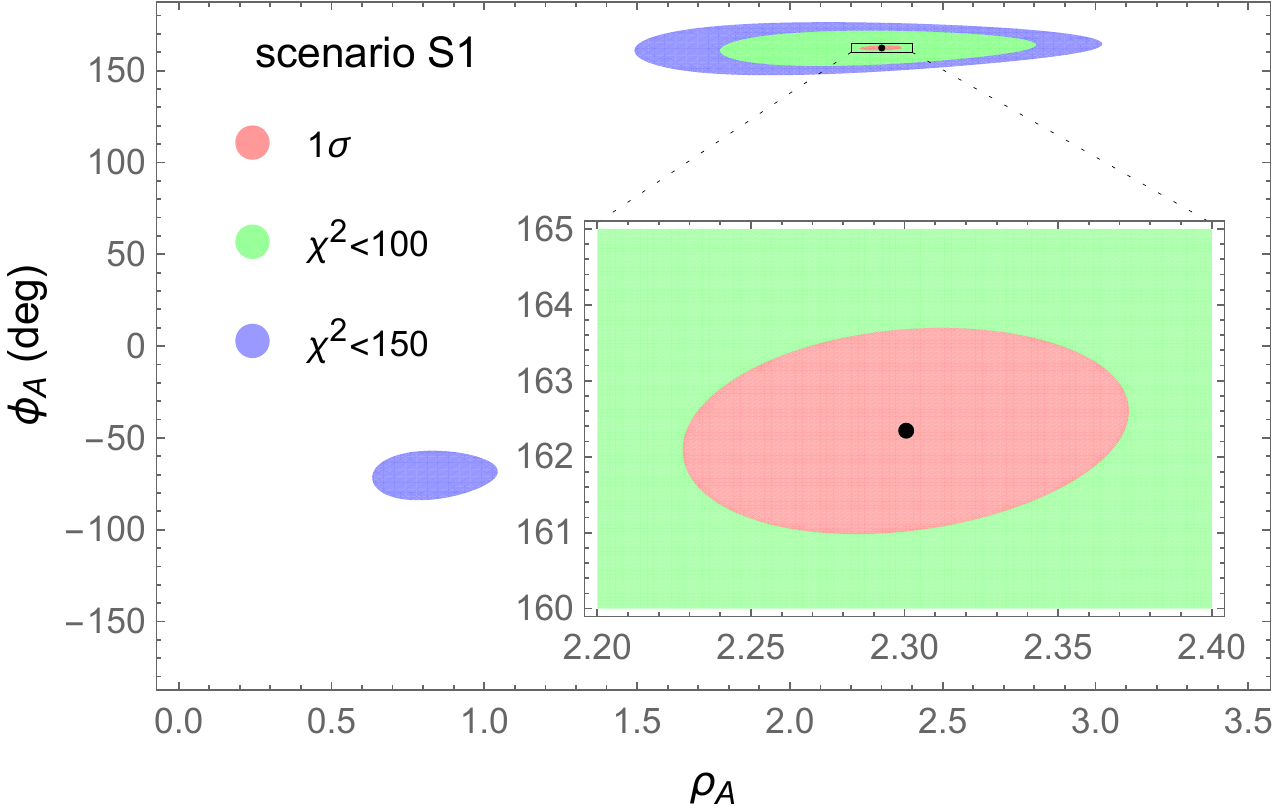}
     \caption{Optimal parameters for Scenario S1 obtained
      from the PDG data.
      In the inset, the oval represents the $1\,{\sigma}$ region
      of the fitted parameters, and the dot denotes the best-fit
      result.}
     \label{fig:s1cov}
     \end{figure}

     The optimal values of the free parameters obtained with
     the PDG data for Scenario S1,
     together with their corresponding correlation coefficients,
     are shown in Table~\ref{tab:para-s1}.
     The best-fit values of the WA parameters ${\rho}_{A}$
     and ${\phi}_{A}$ are illustrated in Fig.~\ref{fig:s1cov}.
     Scenario S1 exhibits two distinctive features.
     The first is that the WA contributions can provide a sizable
     strong phase, which plays an
     essential role in accommodating the observed $CP$ violation.
     The second is that a large value of
     ${\rho}_{A}$ $>$ $2$ is required, echoing the previous results
     reported in Refs. \cite{PhysRevD.90.054019,
     PLB.740.56,PLB.743.444,PhysRevD.91.074026,JPG.43.105004}.
     Such a large ${\rho}_{A}$ poses a substantial
     challenge to the QCDF approach, as it places considerable
     strain on the validity of the power counting scheme.

     \begin{table}[h]
     \caption{Best-fit parameters and their correlations for Scenario S1.}
     \label{tab:para-s1}
     \begin{ruledtabular}
     \begin{tabular}{c c rrr}
     & value & \multicolumn{3}{c}{ correlation matrix }
       \\ \cline{2-2} \cline{3-5}
       ${\chi}^{2}/n_{\rm dof}$
     & $ 52.2 /4 $ & \multicolumn{1}{c}{ $F_{0}^{B{\pi}}$ }
                   & \multicolumn{1}{c}{ ${\rho}_{A}$ }
                   & \multicolumn{1}{c}{ ${\phi}_{A}$ }
       \\ \cline{3-5}
     $ F_{0}^{B{\pi}} $ & $0.251 {\pm} 0.004 $           & $ 1.00$ & $0.01$  & $ -0.18$  \\
     $ {\rho}_{A} $     & $2.30  {\pm} 0.07 $            &         & $ 1.00$ & $  0.19$  \\
     $ {\phi}_{A} $     & $( 162.3 {\pm} 1.4 )^{\circ}$  &         &         & $  1.00$
     \end{tabular}
     \end{ruledtabular}
     \end{table}
     \begin{table}[h]
     \caption{Branching ratios and $CP$ asymmetries for the $B$ ${\to}$
         ${\pi}{\pi}$ decays with the S1 optimal parameters.
         The theoretical uncertainties are
         derived from the errors of the fitted parameters.
         See the text for the definitions of the columns.}
     \label{tab:s1pdg}
     \begin{ruledtabular}
     \begin{tabular}{c ccccc c}
     & \multicolumn{1}{c}{ $ a_{i}       $ }
     & \multicolumn{1}{c}{ $ a_{i,I}^{p} $ }
     & \multicolumn{1}{c}{ $ a_{i}^{p}   $ }
     & \multicolumn{1}{c}{ $ a_{i}     $ $+$ $ b_{i} $ }
     & \multicolumn{1}{c}{ $ a_{i}^{p} $ $+$ $ b_{i} $ }
     & \multicolumn{1}{c}{ PDG } \\ \hline
       $ 10^{6} \,{\times} \, {\cal B}({\pi}^{-}{\pi}^{0}) $
     & $ 5.15 $
     & $ 4.15 $
     & $ 4.32 $
     & $ 5.15 $
     & $ 4.32 {\pm} 0.17 $
     & $ 5.31 {\pm} 0.26 $ \\
       $ 10^{6} \,{\times} \, {\cal B}({\pi}^{+}{\pi}^{-}) $
     & $ 7.11 $
     & $ 7.87 $
     & $ 7.69 $
     & $ 5.61 $
     & $ 6.12 ^{ + 0.32 }_{ - 0.30 } $
     & $ 5.43 {\pm} 0.26 $ \\
       $ 10^{6} \,{\times} \, {\cal B}({\pi}^{0}{\pi}^{0}) $
     & $ 0.20 $
     & $ 0.18 $
     & $ 0.17 $
     & $ 0.70 $
     & $ 0.54 {\pm} 0.06 $
     & $ 1.46 {\pm} 0.19 $ \\
       $ {\cal A}_{CP}({\pi}^{-}{\pi}^{0}) $ &
     & $ -0.004 $
     & $ -0.002 $
     &
     & $ -0.002 {\pm} 0.000 $
     & $ -0.01 {\pm} 0.04 $ \\
       $ {\cal C}_{CP}({\pi}^{+}{\pi}^{-}) $ &
     & $ 0.043 $
     & $ 0.048 $
     & $ -0.375 $
     & $ -0.300 ^{ + 0.051 }_{ - 0.055 } $
     & $ -0.314 {\pm} 0.030 $ \\
       $ {\cal S}_{CP}({\pi}^{+}{\pi}^{-}) $
     & $ -0.420 $
     & $ -0.495 $
     & $ -0.496 $
     & $ -0.592 $
     & $ -0.678 ^{ + 0.038 }_{ - 0.039 } $
     & $ -0.67 {\pm} 0.03 $ \\
       $ {\cal C}_{CP}({\pi}^{0}{\pi}^{0}) $ &
     & $ -0.440 $
     & $ 0.021 $
     & $ 0.442 $
     & $ 0.250 ^{ + 0.064 }_{ - 0.066 } $
     & $ -0.23 {\pm} 0.18 $ \\
       $ {\cal S}_{CP}({\pi}^{0}{\pi}^{0}) $
     & $ 0.994 $
     & $ 0.116 $
     & $ 0.473 $
     & $ 0.884 $
     & $ 0.880 ^{ + 0.031 }_{ - 0.036 } $
     \end{tabular}
     \end{ruledtabular}
     \end{table}
     \begin{figure}[h]
     \includegraphics[width=0.5\textwidth]{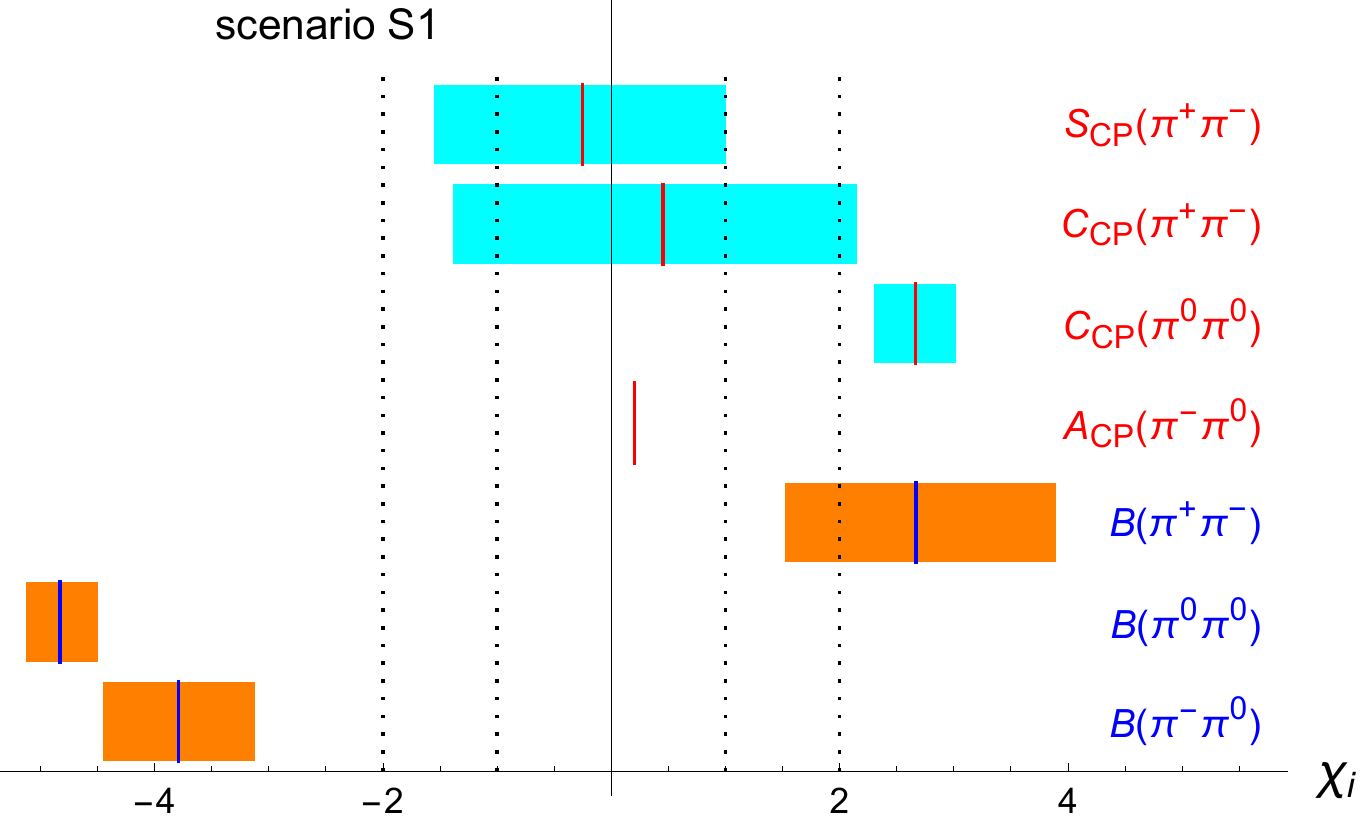}
     \caption{${\chi}_{i}$ distributions with the optimal
      S1 parameters. The vertical bars and horizontal bands
      correspond to the central values and uncertainties of
      the respective observables, respectively.}
     \label{fig:s1xi}
     \end{figure}

     The resulting branching ratios and $CP$ asymmetries,
     calculated using the S1 optimal parameters,
     are presented in Table \ref{tab:s1pdg}.
     The columns labeled ``$a_{i}$'',
     ``$a_{i,I}^{p}$'', ``$a_{i}^{p}$'' correspond
     to the results obtained in the NF scheme with the coefficient
     $a_{i}$ in Eq.(\ref{eq:ai-nf}), and in the QCDF approach
     with the coefficients ``$a_{i,I}^{p}$'' and ``$a_{i}^{p}$''
     in Eq.(\ref{HME-O1-factorizable}), respectively,
     excluding and including the HSS contributions.
     The columns labeled ``$a_{i}$ $+$ $b_{i}$'' and
     ``$a_{i}^{p}$ $+$ $b_{i}$'' denote the results obtained with
     the WA amplitudes ${\cal M}^{a}_{{\pi}{\pi}}$ included.
     The PDG data are also provided for comparison.
     It is clearly seen that the WA contributions have a great impact
     on the branching ratio for the $B$ ${\to}$ ${\pi}^{0}{\pi}^{0}$
     decay.

     The ${\chi}_{i}$ distributions (see Eq.(\ref{eq:chix})),
     which illustrate the discrepancies between the theoretical
     expectations and the experimental measurements,
     are displayed in Fig.~\ref{fig:s1xi},
     with the uncertainties propagated from the fitted parameters.
     As can be seen from Fig.~\ref{fig:s1xi}, the primary
     discrepancies between the theoretical expectations and the
     PDG data originate from the branching ratios, in addition
     to the direct $CP$ asymmetry $ {\cal C}_{CP}({\pi}^{0}{\pi}^{0}) $.
     This tension cannot be easily reconciled
     without going beyond the conventional QCDF framework and
     introducing additional contributions, for instance,
     the soft COME contributions that will be elaborated on
     in the following discussion.

     \subsection{Scenario S2}
     \label{sec0302}
\begin{figure}[h]
     \includegraphics[width=0.4\textwidth]{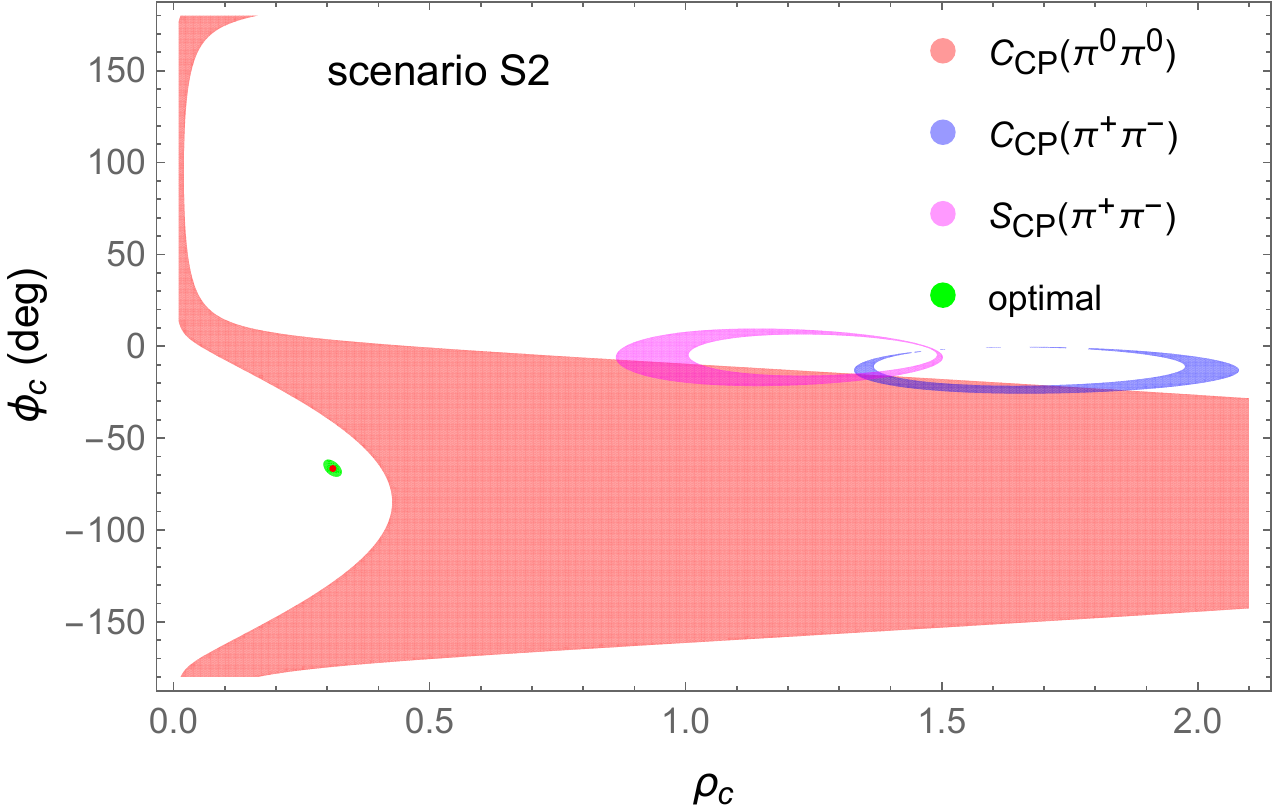}   \qquad
     \includegraphics[width=0.4\textwidth]{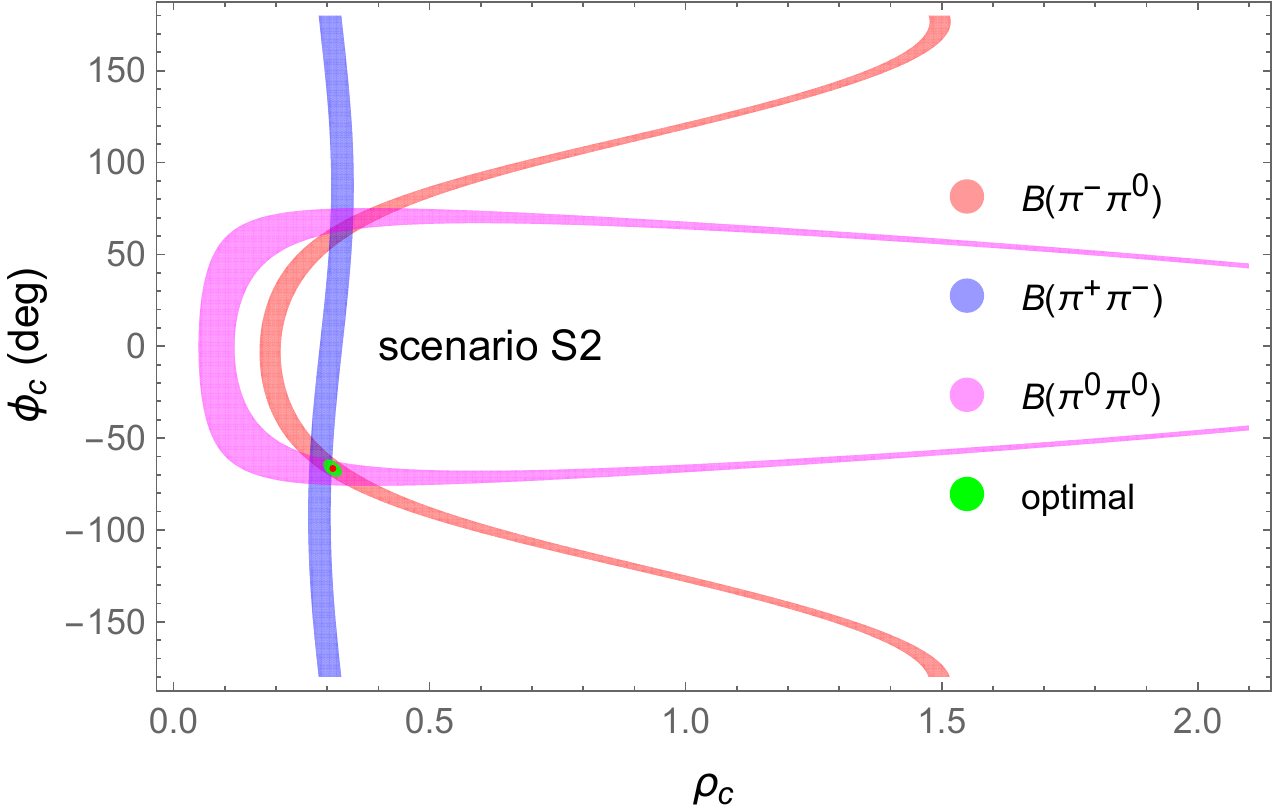}   \\ ~ \\
     \includegraphics[width=0.4\textwidth]{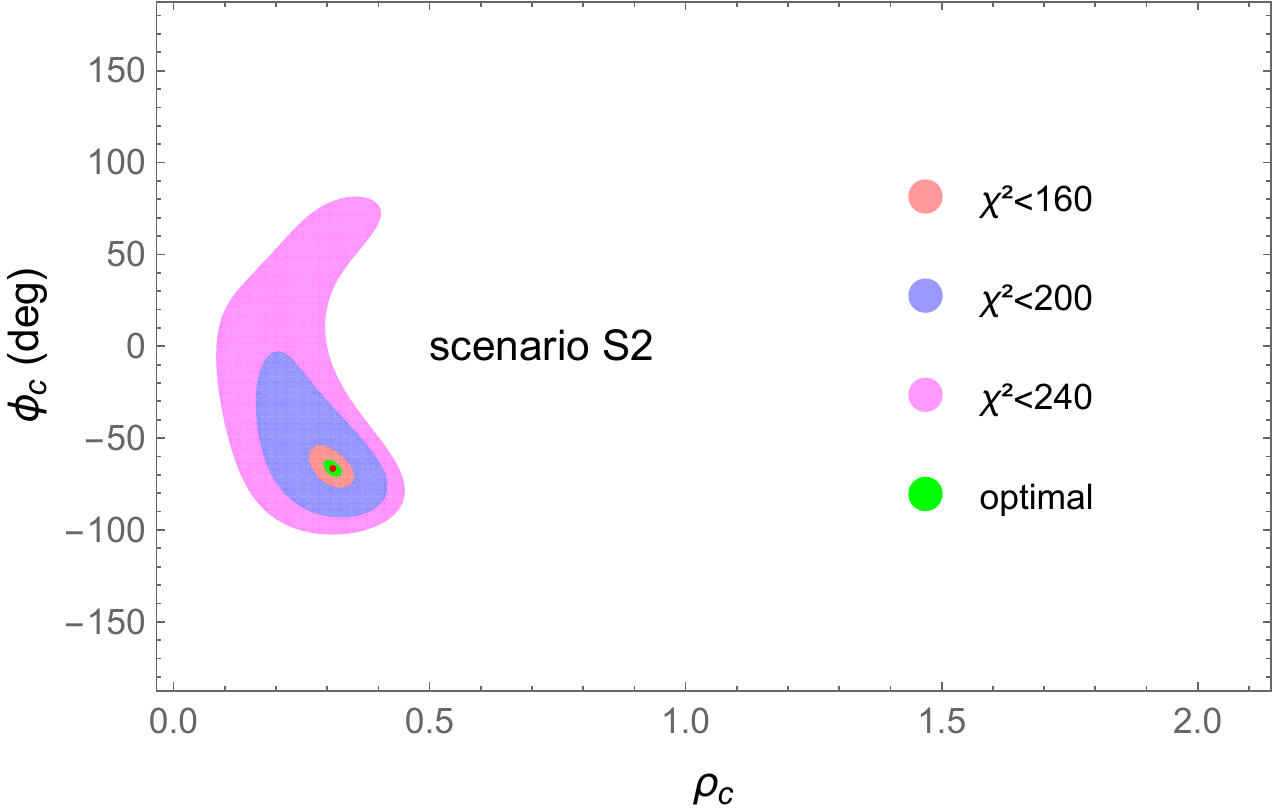}  \qquad
     \includegraphics[width=0.4\textwidth]{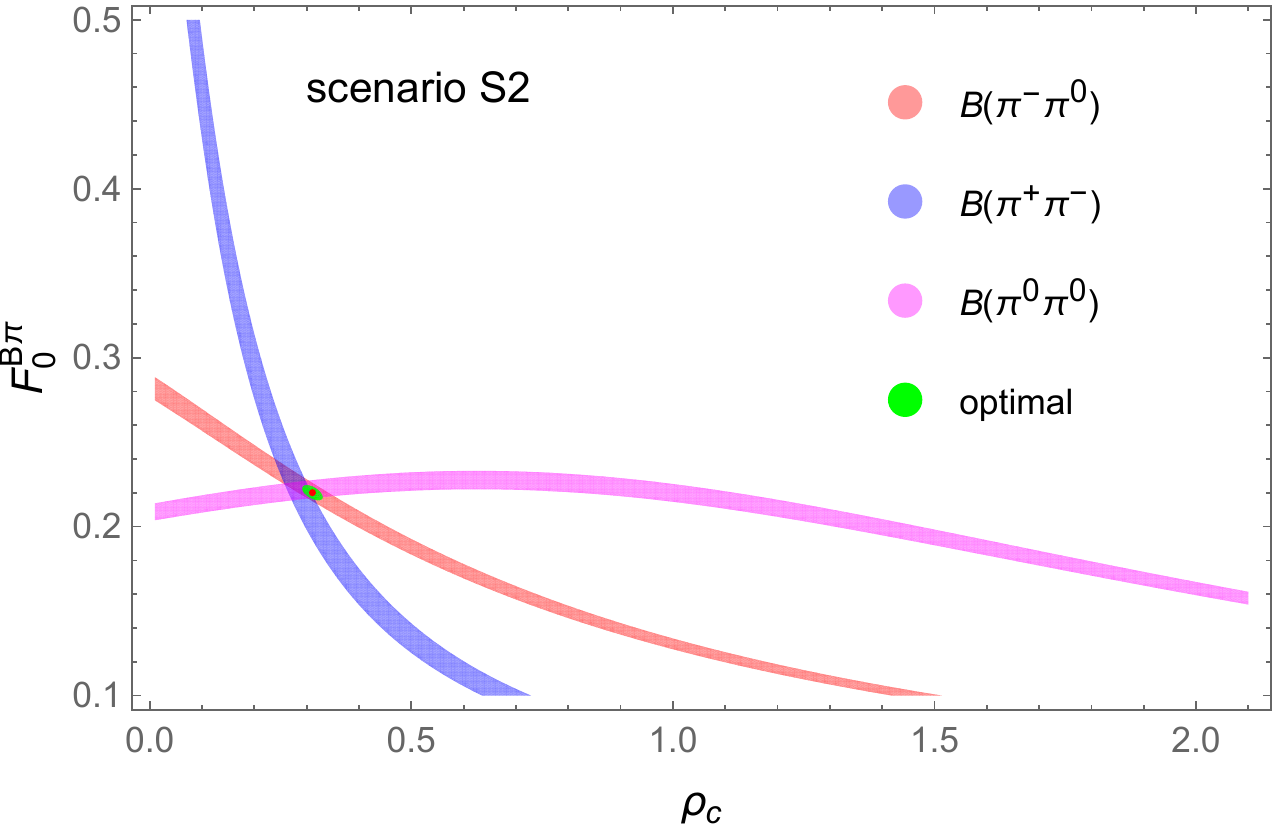}
     \caption{Constraints on the S2 parameters.
        Other legends are the same as those of Fig.~\ref{fig:s1constraints}.}
     \label{fig:s2constraints}
     \end{figure}

     The constraints on the free parameters in Scenario S2
     are derived primarily from branching ratios rather than
     from $CP$ asymmetries, as shown in Fig.~\ref{fig:s2constraints}.
     As is evident from the distribution in the
     (${\rho}_{c}$, ${\phi}_{c}$) plane,
     the simultaneous crossing region of the $CP$ asymmetry bands,
     located near the point ($1.36$, $-17^{\circ}$),
     is far away from the best-fit region.
     This situation is physically unacceptable and indicates that the
     resulting fits cannot provide a consistent and physically
     reasonable interpretation of the $CP$ asymmetry measurements.

     \begin{table}[h]
     \caption{Best-fit parameters and their correlations for Scenario S2.}
     \label{tab:para-s2}
     \begin{ruledtabular}
     \begin{tabular}{c c rrr}
     & value & \multicolumn{3}{c}{ correlation matrix }
       \\ \cline{2-2} \cline{3-5}
       ${\chi}^{2}/n_{\rm dof}$
     & $ 154.1 /4 $ & \multicolumn{1}{c}{ $F_{0}^{B{\pi}}$ }
                   & \multicolumn{1}{c}{ ${\rho}_{c}$ }
                   & \multicolumn{1}{c}{ ${\phi}_{c}$ }
       \\ \cline{3-5}
     $ F_{0}^{B{\pi}} $ & $0.220 {\pm} 0.005 $            & $ 1.00$ & $-0.59$ & $ 0.23$ \\
     $ {\rho}_{c} $     & $0.31 {\pm} 0.02 $              &         & $ 1.00$ & $-0.49$ \\
     $ {\phi}_{c} $     & $( -66.6 {\pm}  4.8 )^{\circ}$  &         &         & $ 1.00$
     \end{tabular}
     \end{ruledtabular}
     \end{table}
     \begin{figure}[h]
     \includegraphics[width=0.5\textwidth]{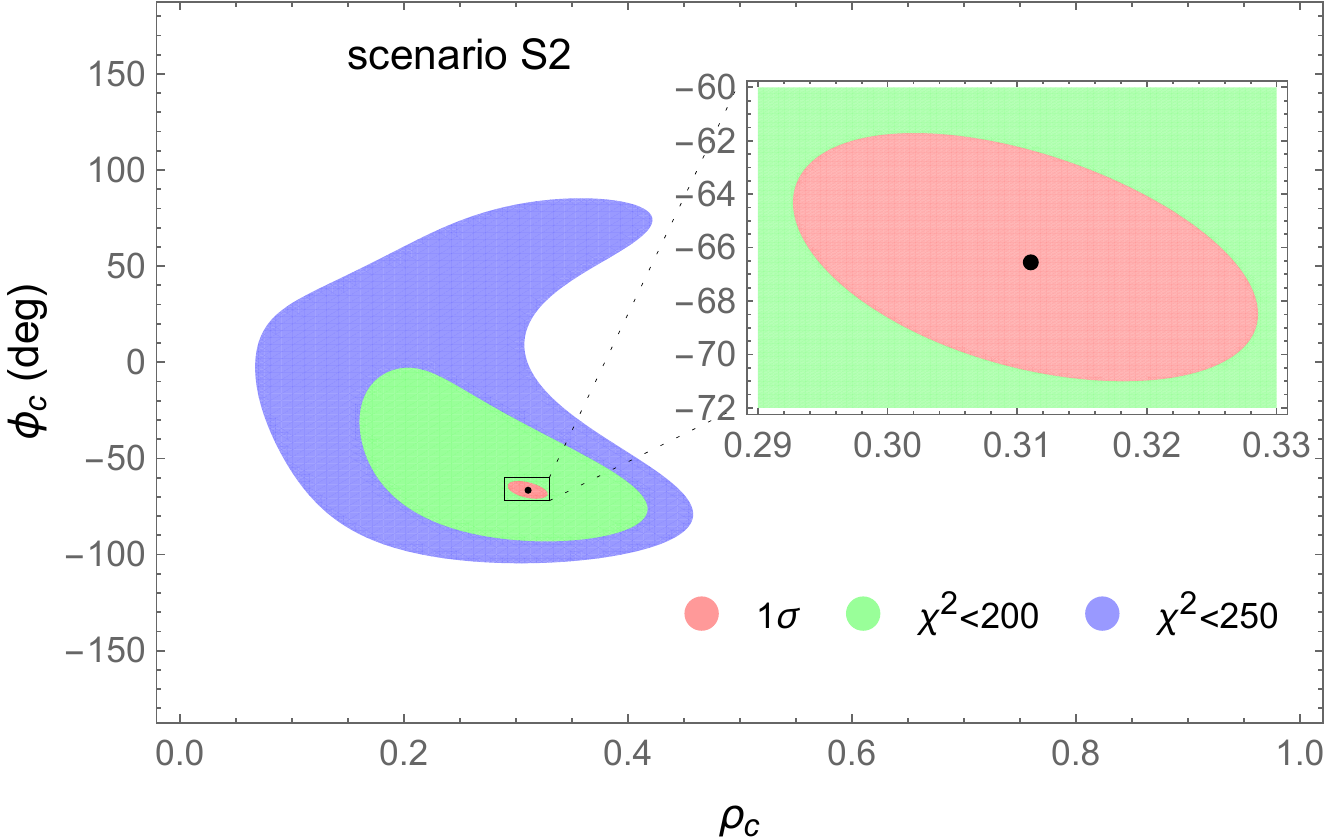}
     \caption{Optimal parameters for Scenario S2 obtained
         from the PDG data.
         The inset shares the same legend as Fig. \ref{fig:s1cov}. }
     \label{fig:s2cov}
     \end{figure}
     \begin{table}[h]
     \caption{Branching ratios and $CP$ asymmetries for the $B$ ${\to}$
         ${\pi}{\pi}$ decays with the S2 optimal parameters.
         Other legends are the same as those of Table~\ref{tab:s1pdg}.}
     \label{tab:s2pdg}
     \begin{ruledtabular}
     \begin{tabular}{c cccc c}
     & \multicolumn{1}{c}{ $ a_{i} $       }
     & \multicolumn{1}{c}{ $ a_{i,I}^{p} $ }
     & \multicolumn{1}{c}{ $ a_{i} $ $+$ $ c_{i} $ }
     & \multicolumn{1}{c}{ $ a_{i,I}^{p} $ $+$ $ c_{i} $ }
     & \multicolumn{1}{c}{ PDG } \\ \hline
      $ 10^{6} \,{\times} \, {\cal B}({\pi}^{-}{\pi}^{0}) $
     & $ 3.97 $
     & $ 3.20 $
     & $ 6.04 $
     & $ 5.29 ^{ + 0.54 }_{ - 0.50 } $
     & $ 5.31 {\pm} 0.26 $ \\
       $ 10^{6} \,{\times} \, {\cal B}({\pi}^{+}{\pi}^{-}) $
     & $ 5.48 $
     & $ 6.06 $
     & $ 4.80 $
     & $ 5.33 {\pm} 0.44 $
     & $ 5.43 {\pm} 0.26 $ \\
       $ 10^{6} \,{\times} \, {\cal B}({\pi}^{0}{\pi}^{0}) $
     & $ 0.16 $
     & $ 0.14 $
     & $ 1.76 $
     & $ 1.67 ^{ + 0.29 }_{ - 0.26 } $
     & $ 1.46 {\pm} 0.19 $ \\
       $ {\cal A}_{CP}({\pi}^{-}{\pi}^{0}) $ &
     & $ -0.004 $
     & $ 0.003 $
     & $ -0.002 $
     & $ -0.01 {\pm} 0.04 $ \\
       $ {\cal C}_{CP}({\pi}^{+}{\pi}^{-}) $ &
     & $ 0.043 $
     & $ -0.012 $
     & $ 0.016 {\pm} 0.003 $
     & $ -0.314 {\pm} 0.030 $ \\
       $ {\cal S}_{CP}({\pi}^{+}{\pi}^{-}) $
     & $ -0.420 $
     & $ -0.495 $
     & $ -0.422 $
     & $ -0.509 {\pm} 0.003 $
     & $ -0.67 {\pm} 0.03 $ \\
       $ {\cal C}_{CP}({\pi}^{0}{\pi}^{0}) $ &
     & $ -0.440 $
     & $ -0.353 $
     & $ -0.512 {\pm} 0.040 $
     & $ -0.23 {\pm} 0.18 $ \\
       $ {\cal S}_{CP}({\pi}^{0}{\pi}^{0}) $
     & $ 0.994 $
     & $ 0.116 $
     & $ 0.066 $
     & $ -0.003 ^{ + 0.054 }_{ - 0.049 } $
     \end{tabular}
     \end{ruledtabular}
     \end{table}
     \begin{figure}[h]
     \includegraphics[width=0.5\textwidth]{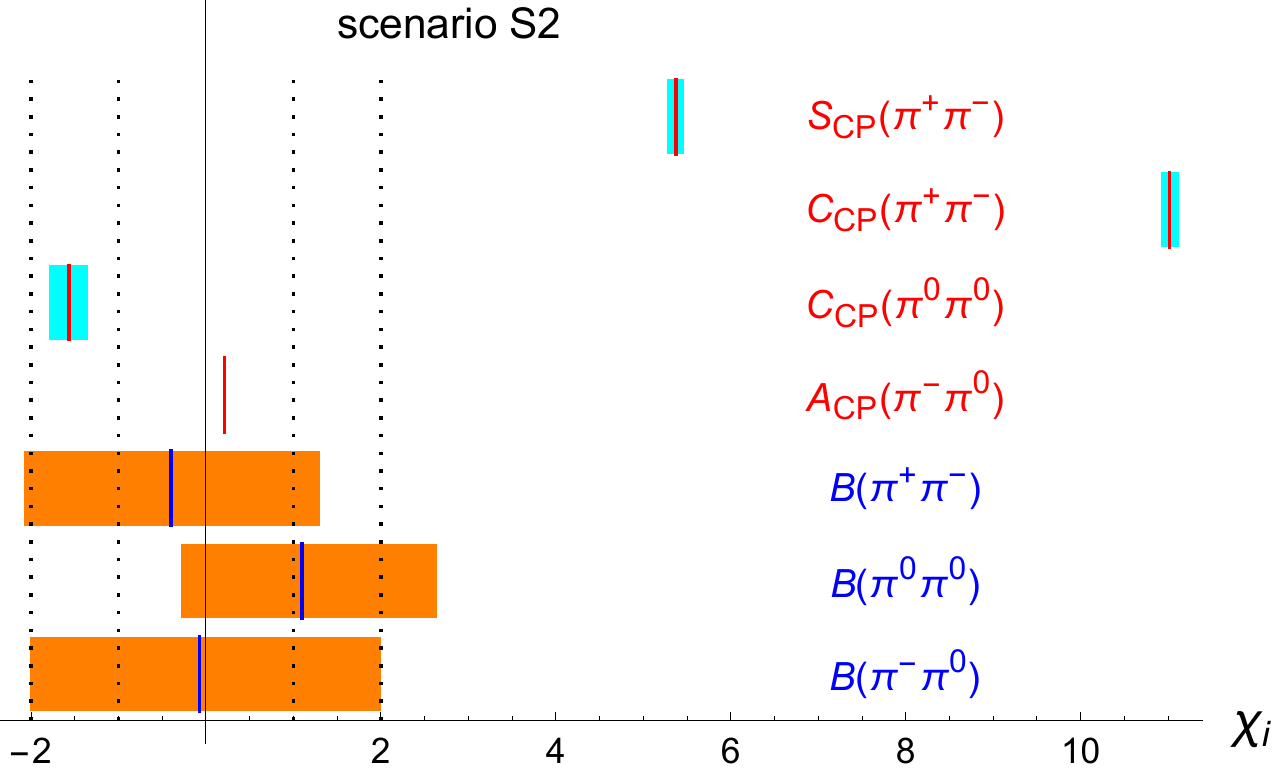}
     \caption{${\chi}_{i}$ distributions for the optimal S2 parameters.
       Other legends are the same as those of Fig.~\ref{fig:s1xi}.}
     \label{fig:s2xi}
     \end{figure}

     The optimal values of the free parameters obtained with the
     PDG data for Scenario S2 are shown in Table \ref{tab:para-s2}.
     The best-fit values of the soft COME parameters ${\rho}_{c}$
     and ${\phi}_{c}$ are illustrated in Fig. \ref{fig:s2cov}.
     Scenario S2 exhibits two distinctive features.
     First, the form factor $F_{0}^{B{\pi}}$ in Scenario S2
     is slightly smaller than its counterpart in Scenario S1.
     This reduction can be largely attributed to the inclusion of
     soft COME contributions in Scenario S2.
     Second, the parameter ${\rho}_{c}$ is found to be around
     $0.3$, suggesting that the soft COME contributions do not
     play a dominant role.
     Furthermore, the value of ${\rho}_{c}$ is
     comparable to that\footnotemark[7]
     \footnotetext[7]{${\rho}_{c}$ corresponds to the parameter
     $d_{1}$ of Eq.(71) in Ref. \cite{PhysRevD.107.013004}.
     It should be noted that the two parameters ${\rho}_{c}$
     and $d_{1}$ are defined within different factorization
     frameworks and should not be directly identified.}
     obtained in the PQCD approach,
     where the contributions from the HSS and WA amplitudes
     are taken into account along with a larger form factor
     $F_{0}^{B{\pi}}$ $=$ $0.27$ \cite{PhysRevD.107.013004}.

     The resulting branching ratios and $CP$ asymmetries,
     calculated using the S2 optimal values of parameters,
     are presented in Table \ref{tab:s2pdg}.
     The columns labeled ``$a_{i}$'', ``$a_{i,I}^{p}$'' correspond
     to the results obtained in the NF scheme with the coefficient
     $a_{i}$ in Eq.(\ref{eq:ai-nf}) and in the QCDF approach
     with the coefficient ``$a_{i,I}^{p}$''
     in Eq.(\ref{HME-O1-factorizable}) excluding
     the HSS contributions, respectively.
     The columns labeled ``$a_{i}$ $+$ $c_{i}$'' and
     ``$a_{i,I}^{p}$ $+$ $c_{i}$'' denote the results obtained
     with the soft COME amplitudes ${\cal M}^{c}_{{\pi}{\pi}}$
     included.
     It is clearly seen that soft COME contributions are remarkably
     significant for the branching ratio of the $B$ ${\to}$
     ${\pi}^{0}{\pi}^{0}$ decay, whereas their impact on the $B$
     ${\to}$ ${\pi}^{+}{\pi}^{-}$ decay is marginal.

     The ${\chi}_{i}$ distributions are displayed in Fig.~\ref{fig:s2xi}.
     It is evident that the primary discrepancies originate from
     the $CP$ asymmetries $ {\cal C}_{CP}({\pi}^{+}{\pi}^{-}) $
     and $ {\cal S}_{CP}({\pi}^{+}{\pi}^{-}) $,
     owing to the absence of the potential strong phases
     associated with the WA contributions.

     \subsection{Scenario S3}
     \label{sec0303}

     \begin{figure}[h]
     \includegraphics[width=0.4\textwidth]{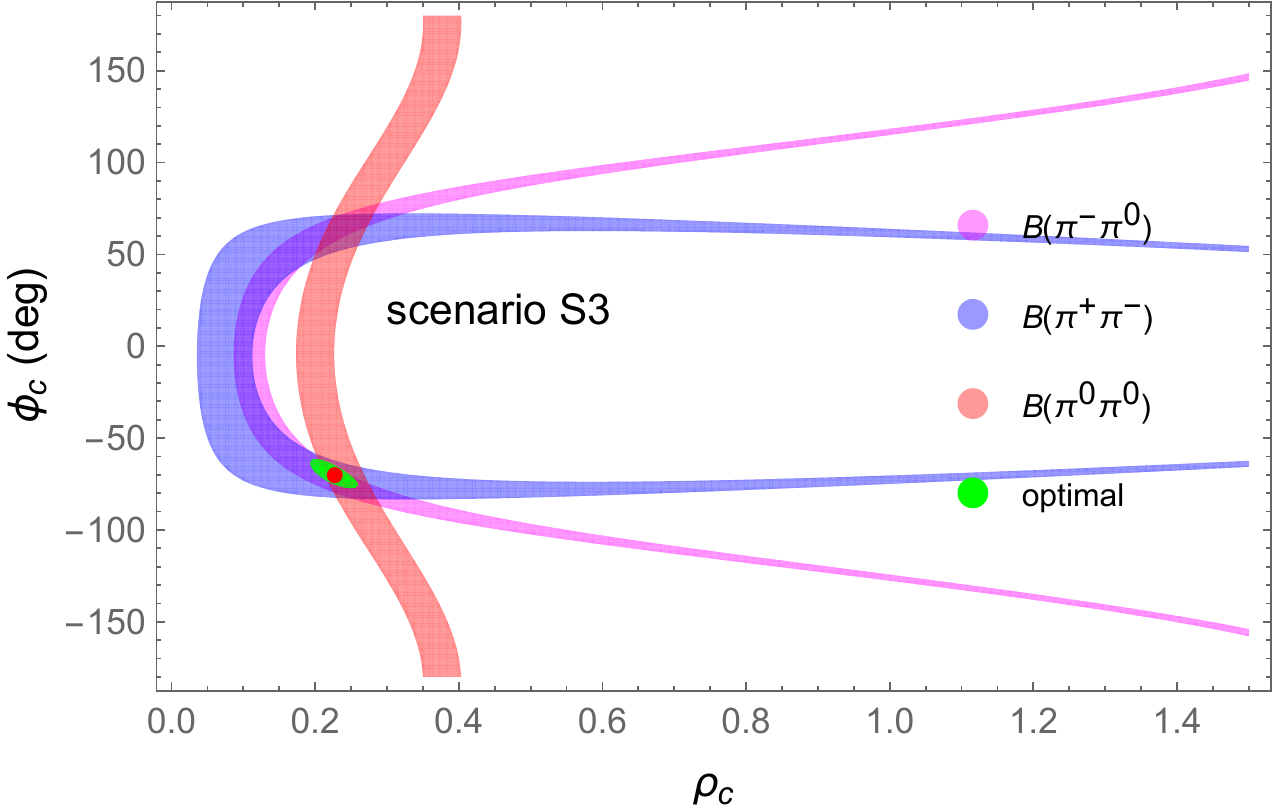} \qquad
     \includegraphics[width=0.4\textwidth]{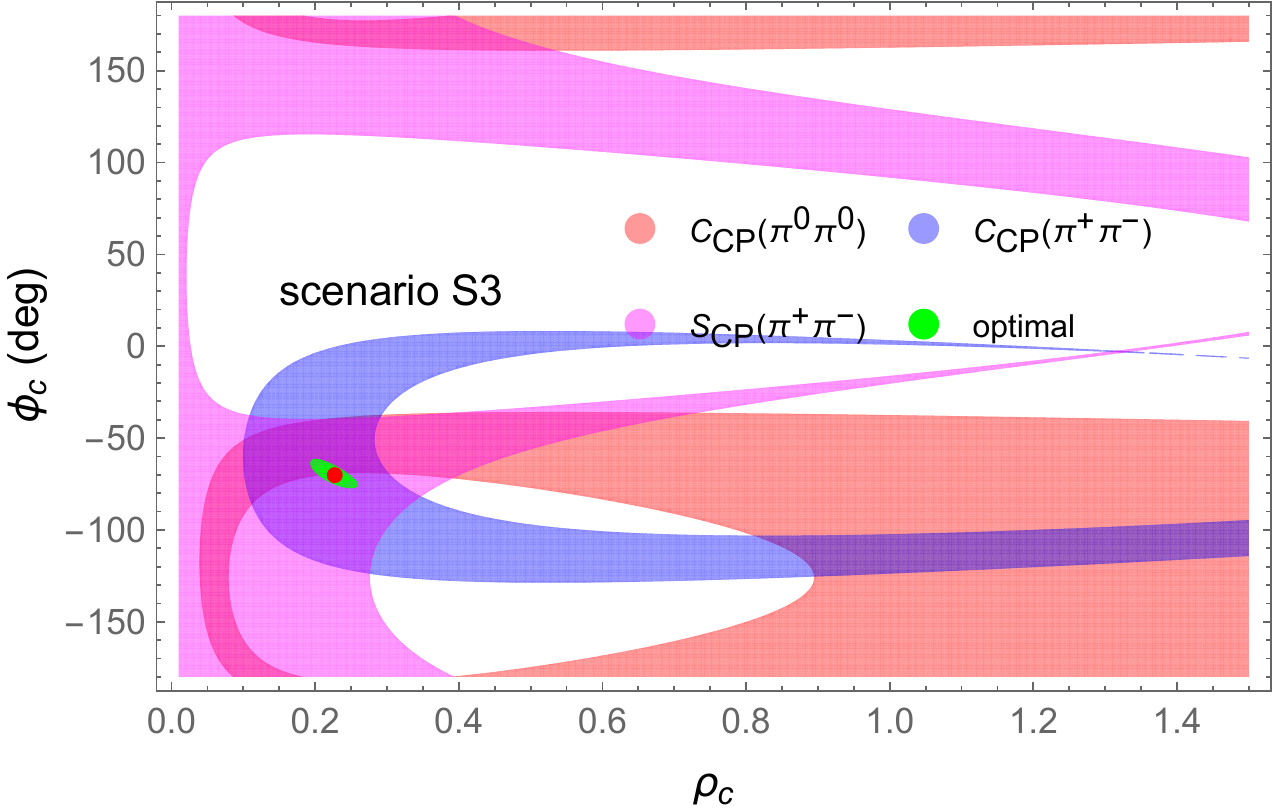} \\ ~ \\
     \includegraphics[width=0.4\textwidth]{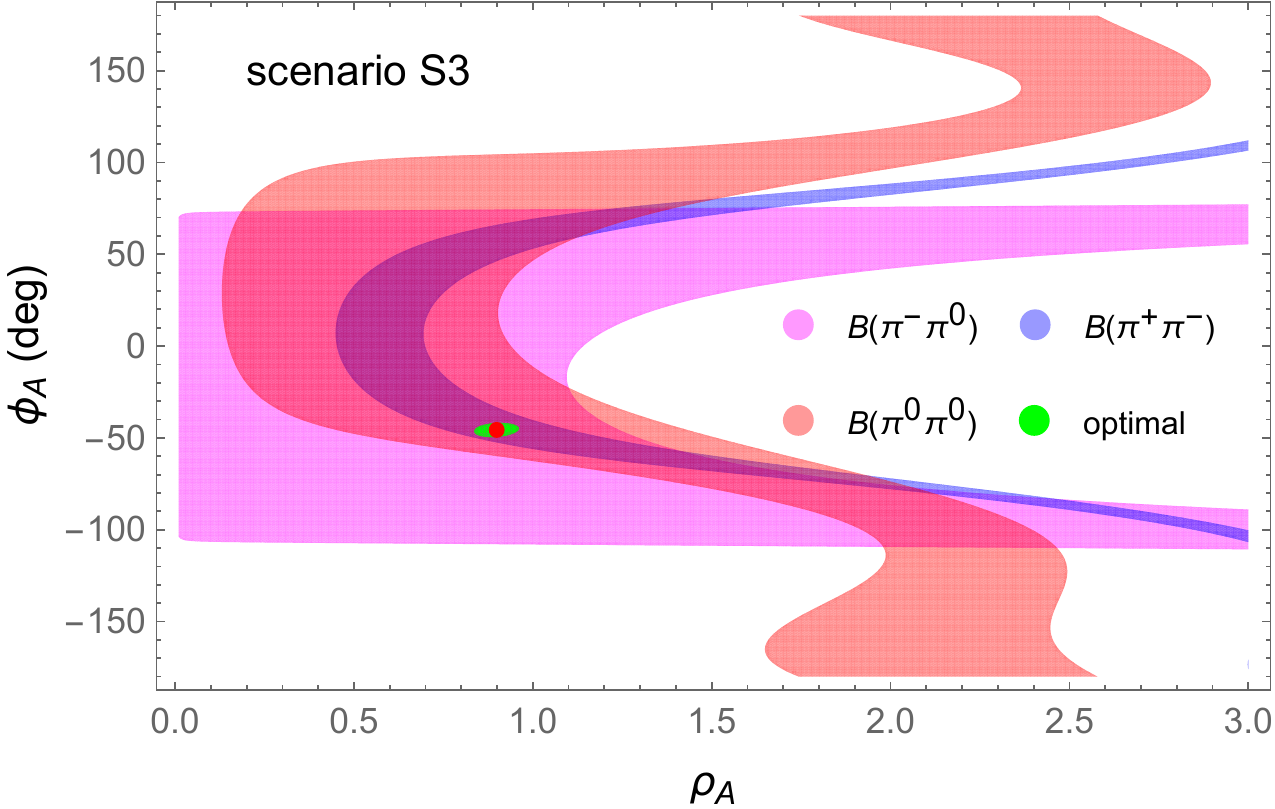} \qquad
     \includegraphics[width=0.4\textwidth]{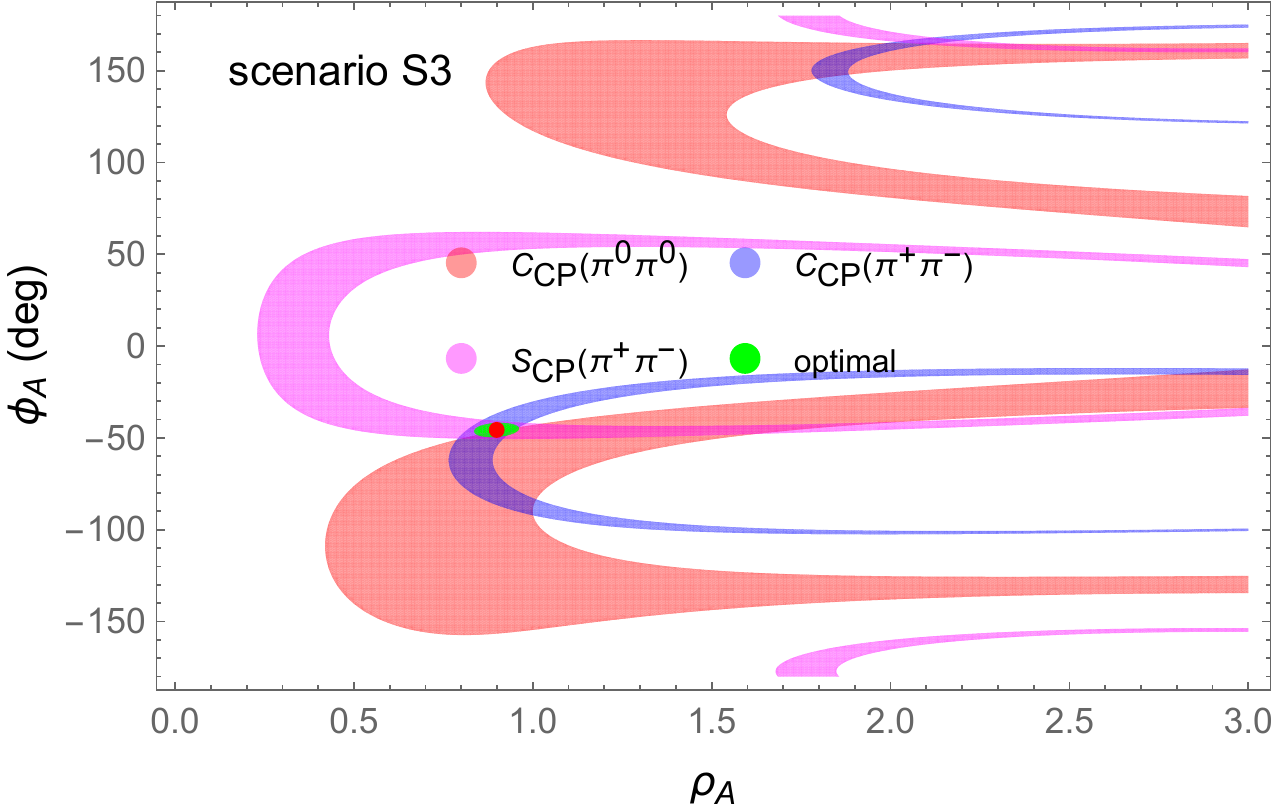} \\ ~ \\
     \includegraphics[width=0.4\textwidth]{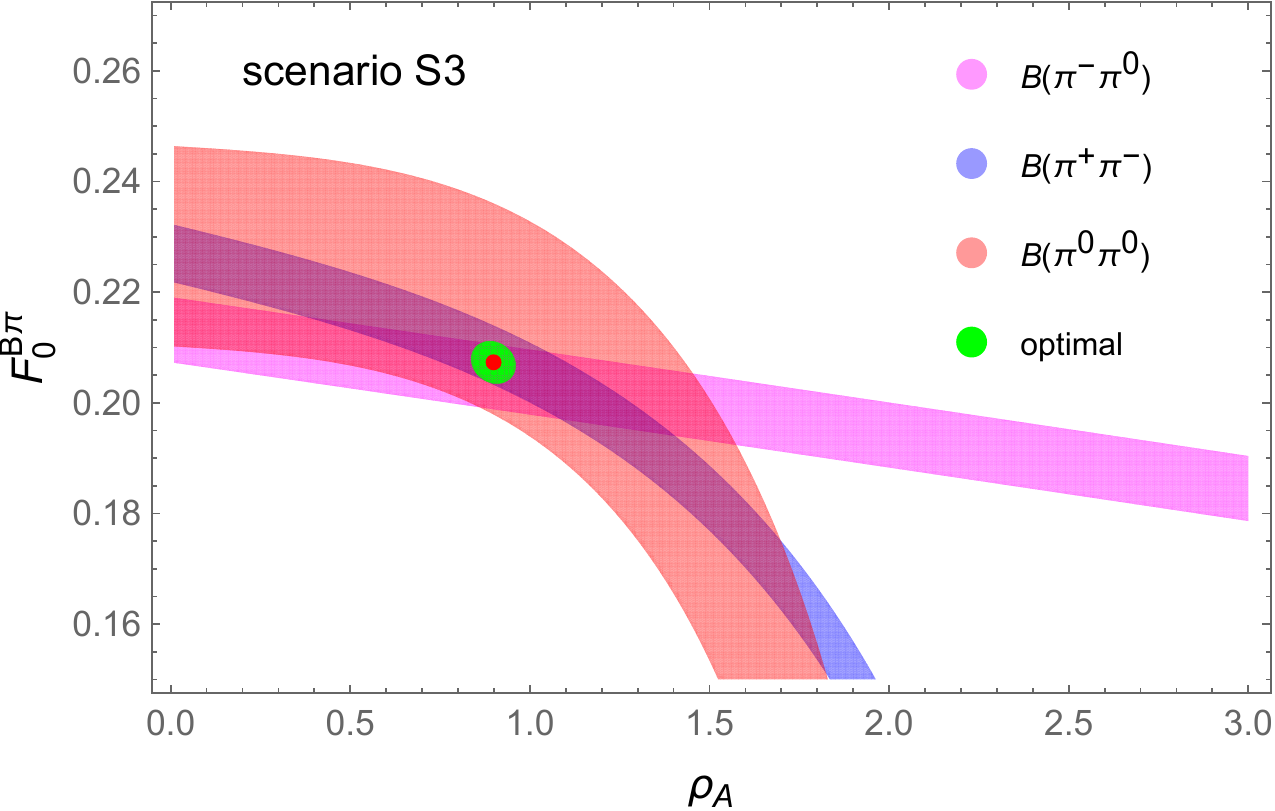} \qquad
     \includegraphics[width=0.4\textwidth]{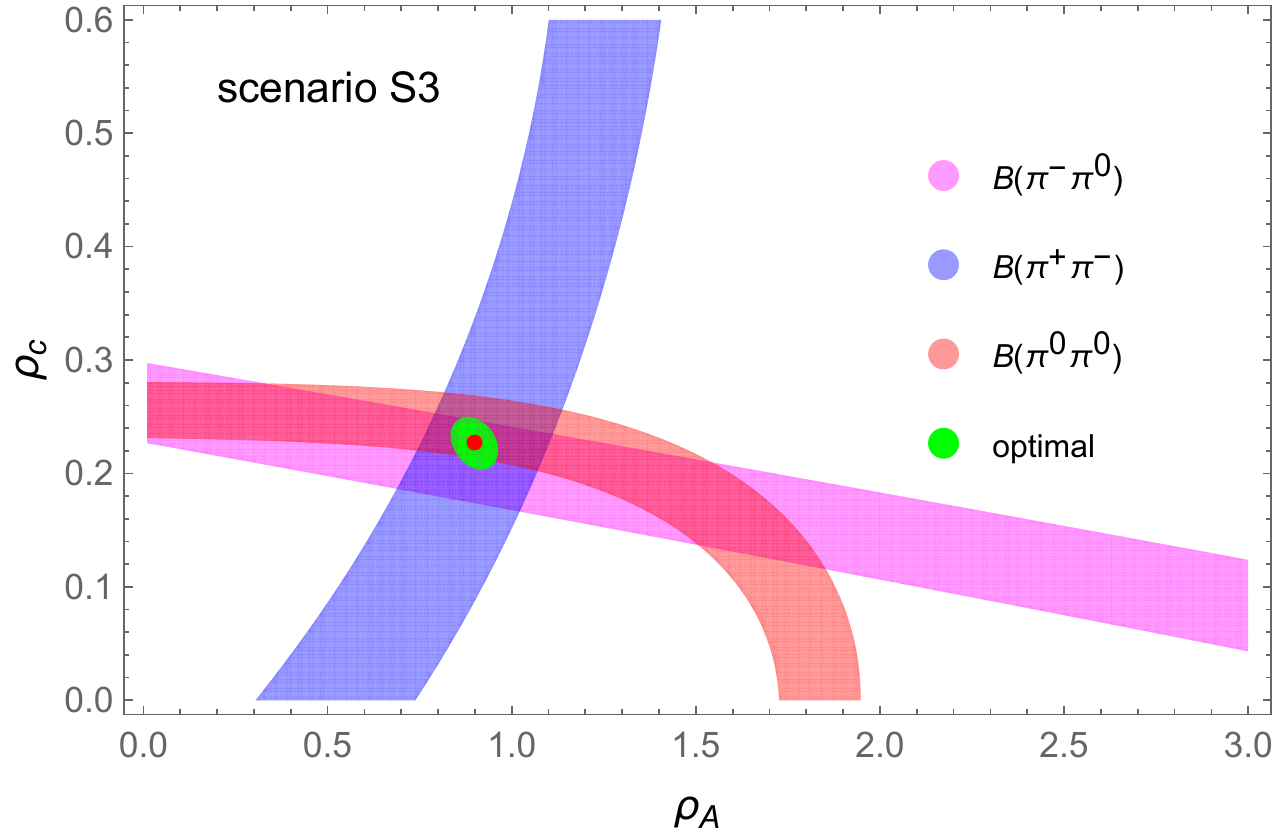}
     \caption{Constraints on the S3 parameters.
        Other legends are the same as those of Fig. \ref{fig:s1constraints}.}
     \label{fig:s3constraints}
     \end{figure}
     \begin{table}[h]
     \caption{Best-fit parameters and their correlations for Scenario S3.}
     \label{tab:para-s3}
     \begin{ruledtabular}
     \begin{tabular}{c c rrrrr}
     & value & \multicolumn{5}{c}{ correlation matrix }
       \\ \cline{2-2} \cline{3-7}
       ${\chi}^{2}/n_{\rm dof}$
     & $ 1.73 /2 $ & \multicolumn{1}{c}{ $F_{0}^{B{\pi}}$ }
                   & \multicolumn{1}{c}{ ${\rho}_{c}$ }
                   & \multicolumn{1}{c}{ ${\phi}_{c}$ }
                   & \multicolumn{1}{c}{ ${\rho}_{A}$ }
                   & \multicolumn{1}{c}{ ${\phi}_{A}$ }
       \\ \cline{3-7}
     $ F_{0}^{B{\pi}} $ & $0.207 {\pm} 0.005 $            & $ 1.00$ & $-0.42$ & $ 0.19$ & $-0.07$ & $-0.44$ \\
     $ {\rho}_{c} $     & $0.23 {\pm} 0.04 $              &         & $ 1.00$ & $-0.77$ & $-0.33$ & $ 0.13$ \\
     $ {\phi}_{c} $     & $ (-70.2 {\pm}  8.4 )^{\circ}$  &         &         & $ 1.00$ & $ 0.28$ & $ 0.04$ \\
     $ {\rho}_{A} $     & $0.90 {\pm} 0.07 $              &         &         &         & $ 1.00$ & $ 0.21$ \\
     $ {\phi}_{A} $     & $( -45.7 {\pm}  4.4 )^{\circ}$  &         &         &         &         & $ 1.00$
     \end{tabular}
     \end{ruledtabular}
     \end{table}
     \begin{figure}[h]
     \includegraphics[width=0.4\textwidth]{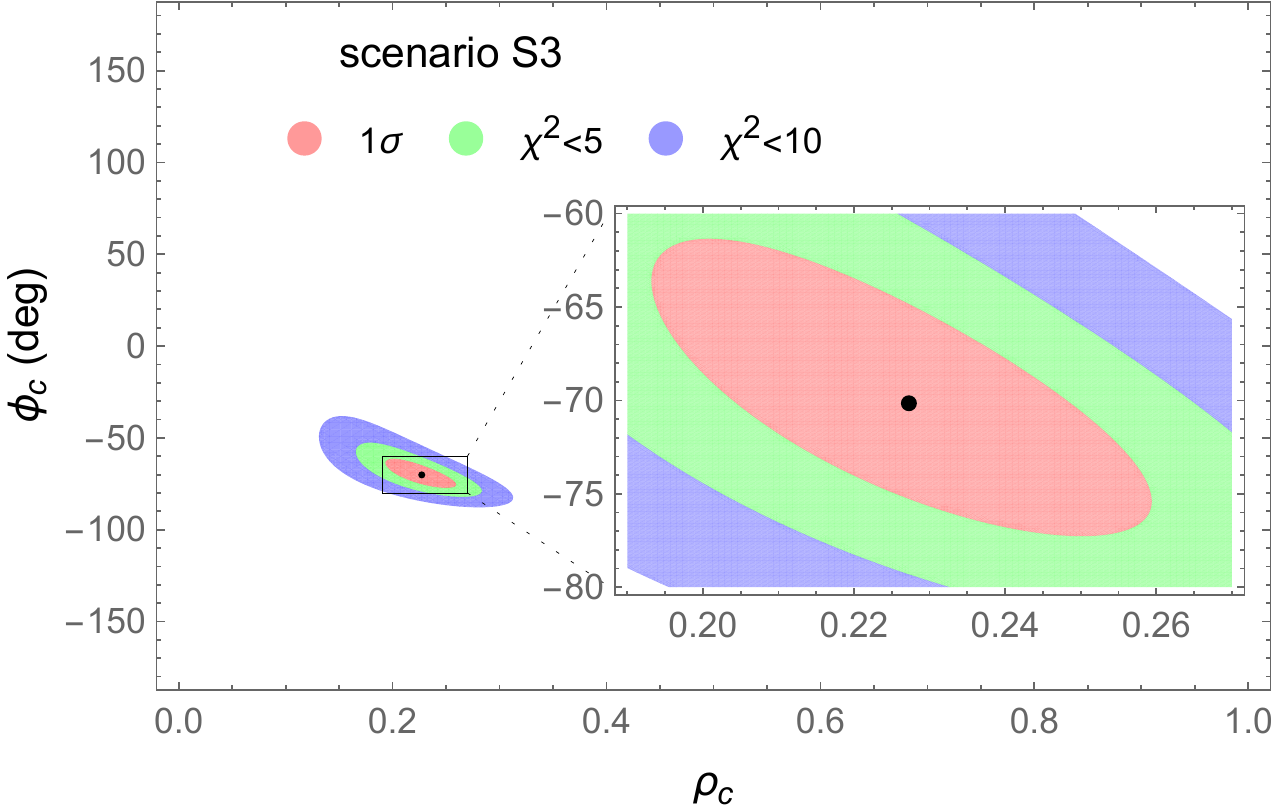} \qquad
     \includegraphics[width=0.4\textwidth]{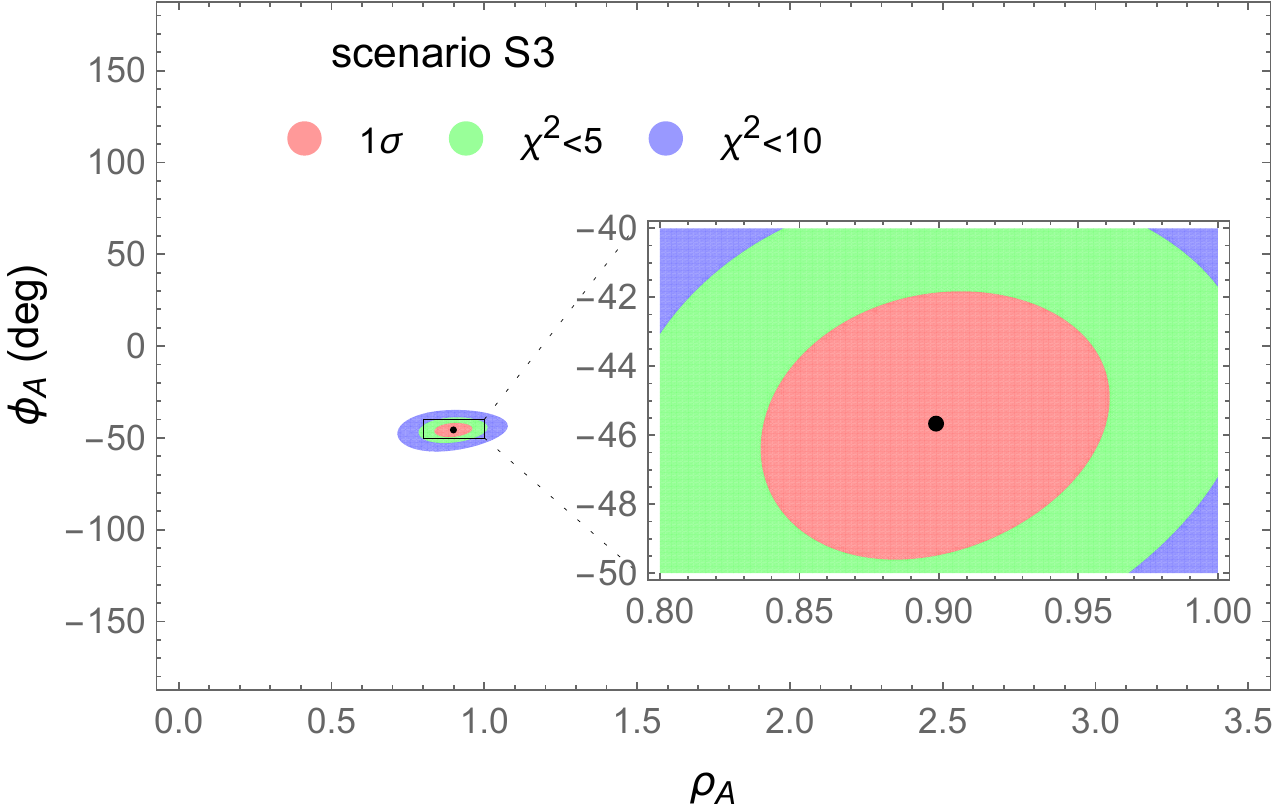}
     \caption{Optimal parameters for Scenario S3 obtained from the PDG data.
        The inset shares the same legend as Fig. \ref{fig:s1cov}}
     \label{fig:s3cov}
     \end{figure}
     \begin{table}[h]
     \caption{Branching ratios and $CP$ asymmetries for the $B$ ${\to}$
         ${\pi}{\pi}$ decays with the S3 optimal parameters.
         Other legends are the same as those of Table~\ref{tab:s1pdg}.}
     \label{tab:s3pdg}
     \begin{ruledtabular}
     \begin{tabular}{c ccccc c}
     & \multicolumn{1}{c}{ $ a_{i} $ }
     & \multicolumn{1}{c}{ $ a_{i}^{p} $ }
     & \multicolumn{1}{c}{ $ a_{i}^{p} $ $+$ $ c_{i} $ }
     & \multicolumn{1}{c}{ $ a_{i}^{p} $ $+$ $ b_{i} $ }
     & \multicolumn{1}{c}{ $ a_{i}^{p} $ $+$ $ b_{i} $ $+$ $ c_{i} $ }
     & \multicolumn{1}{c}{ PDG } \\ \hline
       $ 10^{6} \,{\times} \, {\cal B}({\pi}^{-}{\pi}^{0}) $
     & $ 3.52 $
     & $ 4.10 $
     & $ 5.42 $
     & $ 4.10 $
     & $ 5.42 ^{ + 0.37 }_{ - 0.40 } $
     & $ 5.31 {\pm} 0.26 $ \\
       $ 10^{6} \,{\times} \, {\cal B}({\pi}^{+}{\pi}^{-}) $
     & $ 4.86 $
     & $ 4.32 $
     & $ 3.98 $
     & $ 5.94 $
     & $ 5.36 ^{ + 0.40 }_{ - 0.38 } $
     & $ 5.43 {\pm} 0.26 $ \\
       $ 10^{6} \,{\times} \, {\cal B}({\pi}^{0}{\pi}^{0}) $
     & $ 0.14 $
     & $ 0.40 $
     & $ 1.57 $
     & $ 0.51 $
     & $ 1.36 ^{ + 0.27 }_{ - 0.24 } $
     & $ 1.46 {\pm} 0.19 $ \\
       $ {\cal A}_{CP}({\pi}^{-}{\pi}^{0}) $ &
     & $ -0.005 $
     & $ -0.001 $
     & $ -0.005 $
     & $ -0.001 {\pm} 0.001 $
     & $ -0.01 {\pm} 0.04 $ \\
       $ {\cal C}_{CP}({\pi}^{+}{\pi}^{-}) $ &
     & $ 0.043 $
     & $ 0.017 $
     & $ -0.255 $
     & $ -0.321 ^{ + 0.042 }_{ - 0.044 } $
     & $ -0.314 {\pm} 0.030 $ \\
       $ {\cal S}_{CP}({\pi}^{+}{\pi}^{-}) $
     & $ -0.420 $
     & $ -0.518 $
     & $ -0.528 $
     & $ -0.696 $
     & $ -0.671 ^{ + 0.041 }_{ - 0.044 } $
     & $ -0.67 {\pm} 0.03 $ \\
       $ {\cal C}_{CP}({\pi}^{0}{\pi}^{0}) $ &
     & $ -0.372 $
     & $ -0.440 $
     & $ 0.296 $
     & $ -0.417 ^{ + 0.186 }_{ - 0.150 } $
     & $ -0.23 {\pm} 0.18 $ \\
       $ {\cal S}_{CP}({\pi}^{0}{\pi}^{0}) $
     & $ 0.994 $
     & $ 0.766 $
     & $ 0.150 $
     & $ 0.830 $
     & $ 0.735 ^{ + 0.142 }_{ - 0.167 } $
     \end{tabular}
     \end{ruledtabular}
     \end{table}
     \begin{figure}[h]
     \includegraphics[width=0.5\textwidth]{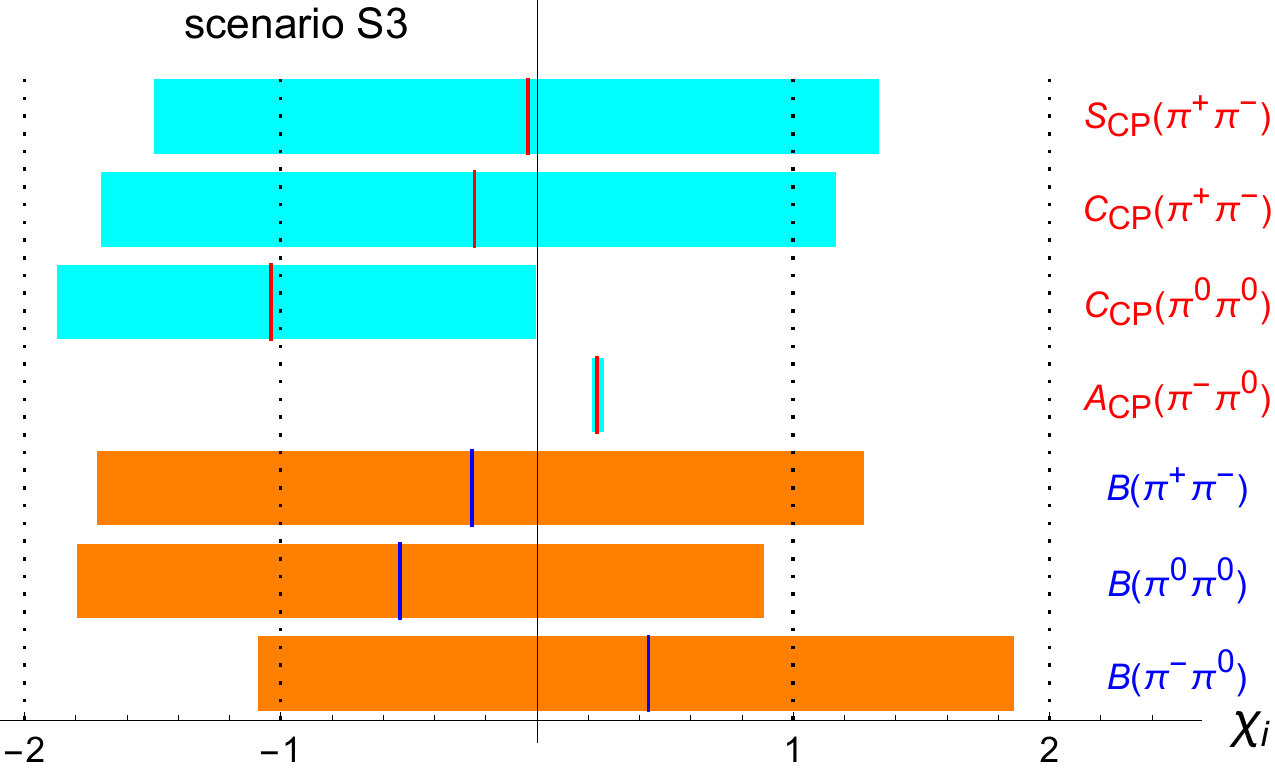}
     \caption{${\chi}_{i}$ distributions for the optimal S3 parameters.
       Other legends are the same as those of Fig.~\ref{fig:s1xi}.}
     \label{fig:s3xi}
     \end{figure}

     The constraints on the free parameters in Scenario S3
     are illustrated in Fig.~\ref{fig:s3constraints}.
     Scenario S3 incorporates more free parameters, thereby
     substantially enhancing the flexibility and effective
     dimensionality of the parameter space.
     In Figs.~\ref{fig:s1constraints} and \ref{fig:s2constraints},
     the parameter regions permitted by the branching ratio constraints
     do not overlap with those allowed by the $CP$ asymmetry constraints.
     In Fig.~\ref{fig:s3constraints}, these previously
     incompatible regions are found to intersect, indicating that
     the internal tensions among the constraints are largely
     alleviated and that the overall compatibility with
     observational data is markedly improved.

     The optimal values of the free parameters obtained from fits
     to the PDG data for Scenario S3 are summarized in
     Table \ref{tab:para-s3}.
     The best-fit values of the parameters are illustrated
     in Fig.~\ref{fig:s3cov}.
     Scenario S3 exhibits several distinctive features.
     (1) The goodness-of-fit, ${\chi}^{2}/n_{\rm dof}$,
     is better in Scenario S3 than in Scenarios S1 and S2.
     (2) Owing to the inclusion of both the WA and soft
     COME contributions, the form factor $F_{0}^{B{\pi}}$ in
     Scenario S3 is smaller than its counterparts in Scenarios
     S1 and S2,
     and also smaller than the typical values adopted in earlier
     QCDF and PQCD analyses \cite{PhysRevD.65.074001,NuclPhysB.675.333,
     NuclPhysB.832.109,PLB.748.422,PhysRevD.80.114008,
     PhysRevD.90.054019,PhysRevD.107.013004,PhysRevD.110.056025}.
     This value is compatible,
     within the quoted uncertainties, with the recent lattice
     QCD determination, $F_{0}^{B{\pi}}$ $=$ $0.183(92)$
     \cite{PhysRevD.106.054502}, as well as with the recent
     light-cone sum rule result, $F_{0}^{B{\pi}}$ $=$
     $0.19(5)$ \cite{JHEP.2023.03.140}.
     (3) The WA parameter ${\rho}_{A}$ is found to be less than $1$.
     This result is consistent with the QCDF conjecture on the
     strength of WA effects and simultaneously underscores the
     importance of soft COME contributions.
     (4)
     The value of the soft COME parameter ${\rho}_{c}$ in
     Scenario S3 is slightly smaller than its counterpart in
     Scenario S2, further reinforcing the importance of WA
     amplitudes.
     The negative correlation between ${\rho}_{c}$ and ${\phi}_{c}$
     is consistent with the pattern found in Scenario S2,
     suggesting that the soft COME parameter space is already
     well constrained by the branching ratio data.
     (5) Owing to the increased number of free parameters
     in Scenario S3, the constraining power of the fit on
     individual parameters is correspondingly weakened,
     leading to larger uncertainties for ${\phi}_{A}$ and
     ${\phi}_{c}$ compared with their counterparts in
     Scenarios S1 and S2, respectively.
     The increased uncertainties for ${\phi}_{A}$ and
     ${\phi}_{c}$ in Scenario S3 reflect the limited
     constraining power of the current experimental data.
     This indicates that, while the combined scenario
     provides a better overall fit, some parameter combinations
     remain loosely constrained.
     Future measurements with improved precision will be essential
     for further pinning down these parameters and for testing the
     viability of Scenario S3.

     The resulting branching ratios and $CP$ asymmetries,
     calculated using the Scenario S3 optimal values of
     parameters, are presented in Table \ref{tab:s3pdg}.
     The columns labeled ``$a_{i}$'', ``$a_{i}^{p}$'' correspond
     to the results obtained in the NF scheme with the
     coefficient $a_{i}$ in Eq.(\ref{eq:ai-nf}) and in the
     QCDF approach with the coefficient ``$a_{i}^{p}$''
     in Eq.(\ref{HME-O1-factorizable}) including
     the HSS contributions, respectively.
     The columns labeled ``$a_{i}^{p}$ $+$ $c_{i}$'' and
     ``$a_{i}^{p}$ $+$ $b_{i}$'' denote the results obtained
     with the soft COME amplitudes ${\cal M}^{c}_{{\pi}{\pi}}$
     and the WA amplitudes ${\cal M}^{a}_{{\pi}{\pi}}$
     included, respectively.
     The columns labeled ``$a_{i}^{p}$ $+$ $b_{i}$ $+$ $c_{i}$''
     denote the preferred results obtained with amplitudes in
     Eq.(\ref{eq:amp-singlet+octet+wa}).
     It is clearly seen that
     soft COME contributions significantly enhance the branching ratio
     of the color-suppressed $B$ ${\to}$ ${\pi}^{0}{\pi}^{0}$ decay,
     and that the WA contributions play a very important role in the
     branching ratio of the color-allowed $B$ ${\to}$ ${\pi}^{+}{\pi}^{-}$
     decay.
     More importantly, for the vast majority of observables,
     including branching ratios and $CP$ asymmetries,
     the Scenario S3 results
     (the ``$a_{i}^{p}$ $+$ $b_{i}$ $+$ $c_{i}$'' column)
     are found to be consistent with the corresponding experimental
     measurements within the quoted uncertainties.
     This overall agreement is further corroborated by the ${\chi}_{i}$
     distributions shown in Fig.~\ref{fig:s3xi},
     which demonstrate that Scenario S3 provides a satisfactory description
     of the data for virtually all $B$ ${\to}$ ${\pi}{\pi}$ decay modes.
     This consistency may offer a possible resolution to the
     so-called ``${\pi}{\pi}$ puzzle''.

     \section{Summary}
     \label{sec04}
     In this paper, we have revisited the $B$ ${\to}$ ${\pi}{\pi}$
     decays within the QCDF framework, with particular emphasis on
     the role of soft COME contributions that are usually
     neglected in the conventional QCDF approach.
     Our primary goal has been to investigate whether the
     inclusion of these soft nonperturbative effects can offer
     a resolution to the long-standing ``${\pi}{\pi}$ puzzle''.

     We have formulated three distinct phenomenological scenarios
     for the fits.
     Scenario S1, which corresponds to the conventional QCDF
     prescription and includes the WA contributions,
     reproduces the previous finding that a large WA parameter,
     ${\rho}_{A}$ $>$ $2$, is required to accommodate the
     experimental data. Such a large value, however, places
     considerable strain on the validity of the power
     counting scheme.
     Scenario S2, which includes soft COME contributions,
     yields a phenomenologically consistent value of
     ${\rho}_{c}$ ${\sim}$ $0.3$, in agreement with the
     PQCD expectation, but fails to account for the observed
     $CP$ asymmetries because the WA-induced strong phases are
     absent.
     The combined scenario, Scenario S3, which incorporates both
     the WA and soft COME amplitudes, provides an excellent overall
     description of almost all observables. The goodness-of-fit for
     the PDG data is significantly improved
     compared with Scenarios S1 and S2, and the theoretical and
     experimental branching ratios and $CP$ asymmetries fall
     within the quoted uncertainties.
     Moreover, the fitted WA parameter in Scenario S3 satisfies
     ${\rho}_{A}$ $<$ $1$, which is consistent with the QCDF
     expectation that WA contributions are power suppressed.

     From the numerical analysis, we draw the following conclusions.
     (1)
     The soft COME contributions play a particularly
     significant role in the color-suppressed $B$ ${\to}$
     ${\pi}^{0}{\pi}^{0}$ decay.
     (2)
     The WA contributions are essential for generating
     the necessary strong phases to describe the $CP$ asymmetries.
     (3)
     Only when both types of contributions are included
     simultaneously can the QCDF framework accommodate almost all
     $B$ ${\to}$ ${\pi}{\pi}$ observables in a theoretically
     consistent and empirically satisfactory manner.

     In summary, our results indicate that soft COME contributions,
     overlooked in the conventional QCDF treatment, may serve as
     a crucial ingredient for resolving the ``${\pi}{\pi}$ puzzle''.
     The soft COME amplitudes naturally explain the observed hierarchy
     among the branching ratios without invoking unnaturally large
     WA parameters.
     This work therefore provides a possible way to reconcile the
     long-standing discrepancies between theory and experiment
     within the SM, and calls for further theoretical scrutiny of
     soft COME contributions from first-principles QCD calculations.
     Future high-precision measurements will be essential for
     further testing the scenario proposed in this work.

     \section*{Acknowledgments}
     This work is supported by the National Natural Science Foundation of China (Grant No. 12275068),
     the National Key R\&D Program of China (Grant No. 2023YFA1606000), and
     the Natural Science Foundation of Henan Province (Grant Nos. 262300421308, 252300421491).



\begin{thebibliography}{99}
     \bibitem{PDG.2026}
         F. Takahashi, K. Hikasa, Y. Sumino {\it et al.} (Particle Data Group),
         Review of particle physics,
         \href{https://doi.org/10.1142/S0217751X26300115}
              {Int. J. Mod. Phys. A 41, 2630011 (2026).}

     \bibitem{PhysRevD.41.1522}
         C. Dib, I. Dunietz, F. Gilman, Y. Nir,
         Standard-model predictions for $CP$ violation in $B^{0}$-meson decay,
         \href{https://doi.org/10.1103/PhysRevD.41.1522}
              {Phys. Rev. D 41, 1522 (1990).}
     \bibitem{PhysRevLett.65.3381}
         M. Gronau, D. London,
         Isospin analysis of $CP$ asymmetries in $B$ decays,
         \href{https://doi.org/10.1103/PhysRevLett.65.3381}
              {Phys. Rev. Lett. 65, 3381 (1990).}
     \bibitem{PhysRevD.44.1454}
         H. Lipkin, Y. Nir, H. Quinn, A. Snyder,
         Penguin trapping with isospin analysis and $CP$ asymmetries in $B$ decays,
         \href{https://doi.org/10.1103/PhysRevD.44.1454}
              {Phys. Rev. D 44, 1454 (1991).}

     \bibitem{PRL.10.531}
         N. Cabibbo,
         Unitary symmetry and leptonic decays,
         \href{https://doi.org/10.1103/PhysRevLett.10.531}
              {Phys. Rev. Lett. 10, 531 (1963).}
     \bibitem{PTP.49.652}
         M. Kobayashi, T. Maskawa,
         $CP$-violation in the renormalizable theory of weak interaction,
         \href{https://doi.org/10.1143/PTP.49.652}
              {Prog. Theor. Phys. 49, 652 (1973).}

     \bibitem{PhysRevLett.83.1914}
         M. Beneke, G. Buchalla, M. Neubert, C. Sachrajda,
         QCD factorization for $B$ ${\to}$ ${\pi}{\pi}$ decays: strong phases
         and $CP$ violation in the heavy quark limit,
         \href{https://doi.org/10.1103/PhysRevLett.83.1914}
              {Phys. Rev. Lett. 83, 1914 (1999).}
     \bibitem{NuclPhysB.591.313}
         M. Beneke, G. Buchalla, M. Neubert, C. Sachrajda,
         QCD factorization for exclusive nonleptonic $B$ meson decays:
         General arguments and the case of heavy light final states,
         \href{https://doi.org/10.1016/S0550-3213(00)00559-9}
              {Nucl. Phys. B 591, 313 (2000).}
     \bibitem{PhysLettB.488.46}
         D. Du, D. Yang, G. Zhu,
         Analysis of the decays $B$ ${\to}$ ${\pi}{\pi}$ and ${\pi}K$
         with QCD factorization in the heavy quark limit,
         \href{https://doi.org/10.1016/S0370-2693(00)00854-6}
              {Phys. Lett. B 488, 46 (2000).}
     \bibitem{PhysLettB.509.263}
         D. Du, D. Yang, G. Zhu,
         Infrared divergence and twist-3 distribution amplitudes in QCD factorization for $B$ ${\to}$ $PP$,
         \href{https://doi.org/10.1016/S0370-2693(01)00398-7}
              {Phys. Lett. B 509, 263 (2001).}
     \bibitem{PhysRevD.64.014036}
         D. Du, D. Yang, G. Zhu,
         QCD factorization for $B$ ${\to}$ $PP$,
         \href{https://doi.org/10.1103/PhysRevD.64.014036}
              {Phys. Rev. D 64, 014036 (2001).}
     \bibitem{NuclPhysB.606.245}
         M. Beneke, G. Buchalla, M. Neubert, C. Sachrajda,
         QCD factorization in $B$ ${\to}$ ${\pi}K$, ${\pi}{\pi}$
         decays and extraction of Wolfenstein parameters,
         \href{https://doi.org/10.1016/S0550-3213(01)00251-6}
              {Nucl. Phys. B 606, 245 (2001).}
     \bibitem{PhysRevLett.74.4388}
         H. Li, H. Yu,
         Extraction of $V_{ub}$ from the decay $B$ ${\to}$ ${\pi}{\ell}{\nu}$,
         \href{https://doi.org/10.1103/PhysRevLett.74.4388}
              {Phys. Rev. Lett. 74, 4388 (1995).}
     \bibitem{PhysLettB.348.597}
         H. Li,
         Study of the decay $B$ ${\to}$ ${\pi}{\pi}$ via perturbative QCD,
         \href{https://doi.org/10.1016/0370-2693(95)00174-J}
              {Phys. Lett. B 348, 597 (1995).}
     \bibitem{PhysLettB.504.6}
         Y. Keum, H. Li, A. Sanda,
         Fat penguins and imaginary penguins in perturbative QCD,
         \href{https://doi.org/10.1016/S0370-2693(01)00247-7}
              {Phys. Lett. B 504, 6 (2001).}
     \bibitem{PhysRevD.63.054008}
         Y. Keum, H. Li, A. Sanda,
         Penguin enhancement and $B$ ${\to}$ $K{\pi}$ decays in perturbative QCD,
         \href{https://doi.org/10.1103/PhysRevD.63.054008}
              {Phys. Rev. D 63, 054008 (2001).}
     \bibitem{PhysRevD.63.074006}
         Y. Keum, H. Li,
         Nonleptonic charmless $B$ decays: factorization versus perturbative QCD,
         \href{https://doi.org/10.1103/PhysRevD.63.074006}
              {Phys. Rev. D 63, 074006 (2001).}
     \bibitem{PhysRevD.63.074009}
         C. L\"{u}, K. Ukai, M. Yang,
         Branching ratio and $CP$ violation of $B$ ${\to}$ ${\pi}{\pi}$
         decays in the perturbative QCD approach,
         \href{https://doi.org/10.1103/PhysRevD.63.074009}
              {Phys. Rev. D 63, 074009 (2001).}
     \bibitem{EPJC.23.275}
         C. L\"{u}, M. Yang,
         $B$ ${\to}$ ${\pi}{\rho}$, ${\pi}{\omega}$ decays in perturbative QCD approach,
         \href{https://doi.org/10.1007/s100520100878}
              {Eur. Phys. J. C 23, 275 (2002).}
     \bibitem{PhysLettB.555.197}
         H. Li, K. Ukai,
         Threshold resummation for nonleptonic $B$ meson decays,
         \href{https://doi.org/10.1016/S0370-2693(03)00049-2}
              {Phys. Lett. B 555, 197 (2003).}

     \bibitem{PhysRevD.87.031103}
         Y. Duh, T. Wu, P. Chang {\it et al.} (Belle Collaboration),
         Measurements of branching fractions and direct $CP$ asymmetries for
         $B$ ${\to}$ $K{\pi}$, $B$ ${\to}$ ${\pi}{\pi}$ and $B$ ${\to}$ $KK$ decays,
         \href{https://doi.org/10.1103/PhysRevD.87.031103}
              {Phys. Rev. D 87, 031103 (2013).}
     \bibitem{PhysRevD.96.032007}
         T. Julius, M. Sevior, G. Mohanty  {\it et al.} (Belle Collaboration),
         Measurement of the branching fraction and $CP$ asymmetry in
         $B^{0}$ ${\to}$ ${\pi}^{0}{\pi}^{0}$ decays, and an improved
         constraint on ${\phi}_{2}$,
         \href{https://doi.org/10.1103/PhysRevD.96.032007}
              {Phys. Rev. D 96, 032007 (2017).}
     \bibitem{PhysRevD.88.092003}
         J. Dalseno, K. Prothmann, C. Kiesling {\it et al.} (Belle Collaboration),
         Measurement of the $CP$ violation parameters in $B^{0}$ ${\to}$ ${\pi}^{+}{\pi}^{-}$ decays,
         \href{https://doi.org/10.1103/PhysRevD.88.092003}
              {Phys. Rev. D 88, 092003 (2013).}
     \bibitem{PhysRevD.76.091102}
         B. Aubert, M. Bona, D. Boutigny {\it et al.} (BaBar Collaboration),
         Study of $B^{0}$ ${\to}$ ${\pi}^{0}{\pi}^{0}$,
         $B^{\pm}$ ${\to}$ ${\pi}^{\pm}{\pi}^{0}$,
         and $B^{\pm}$ ${\to}$ $K^{\pm}{\pi}^{0}$
         decays, and isospin analysis of $B$ ${\to}$ ${\pi}{\pi}$ decays,
         \href{https://doi.org/10.1103/PhysRevD.76.091102}
              {Phys. Rev. D 76, 091102 (2007).}
     \bibitem{PhysRevD.87.052009}
         J. Lees, V. Poireau, V. Tisserand {\it et al.} (BaBar Collaboration),
         Measurement of $CP$ asymmetries and branching fractions in charmless
         two-body $B$-meson decays to pions and kaons,
         \href{https://doi.org/10.1103/PhysRevD.87.052009}
              {Phys. Rev. D 87, 052009 (2013).}
     \bibitem{PhysRevD.75.012008}
         B. Aubert, R. Barate, M. Bona {\it et al.} (BaBar Collaboration),
         Improved measurements of the branching fractions for
         $B^{0}$ ${\to}$ ${\pi}^{+}{\pi}^{-}$ and
         $B^{0}$ ${\to}$ $K^{+}{\pi}^{-}$, and a search for
         $B^{0}$ ${\to}$ $K^{+}K^{-}$,
         \href{https://doi.org/10.1103/PhysRevD.75.012008}
              {Phys. Rev. D 75, 012008 (2007).}
     \bibitem{PhysRevD.109.012001}
         I. Adachi, L. Aggarwal, H. Ahmed {\it et al.} (Belle II Collaboration),
         Measurement of branching fractions and direct $CP$ asymmetries for
         $B$ ${\to}$ $K{\pi}$ and $B$ ${\to}$ ${\pi}{\pi}$ decays at Belle II,
         \href{https://doi.org/10.1103/PhysRevD.109.012001}
              {Phys. Rev. D 109, 012001 (2024).}
     \bibitem{PhysRevD.111.L071102}
         I. Adachi, L. Aggarwal, H. Ahmed {\it et al.} (Belle II Collaboration),
         Measurement of the branching fraction and $CP$-violating asymmetry of
         the decay $B^{0}$ ${\to}$ ${\pi}^{0} {\pi}^{0}$ using $387$ million
         ${\Upsilon}(4S)$ decays in Belle II data,
         \href{https://doi.org/10.1103/PhysRevD.111.L071102}
              {Phys. Rev. D 111, L071102 (2025).}
     \bibitem{2606.20797}
         M. Abumusabh, I. Adachi, K. Adamczyk {\it et al.} (Belle II Collaboration),
         Measurement of mixing-induced $CP$ violating in the decay $B^{0}$ ${\to}$ ${\pi}^{0} {\pi}^{0}$,
         \href{https://doi.org/10.48550/arXiv.2606.20797}
              {arXiv.2606.20797.}

     \bibitem{PhysRevD.65.074001}
         D. Du, H. Gong, J. Sun, D. Yang, G. Zhu,
         Phenomenological analysis of $B$ ${\to}$ $PP$ decays with QCD factorization,
         \href{https://doi.org/10.1103/PhysRevD.65.074001}
              {Phys. Rev. D 65, 074001 (2002).}
     \bibitem{NuclPhysB.675.333}
         M. Beneke, M. Neubert,
         QCD factorization for $B$ ${\to}$ $PP$ and $B$ ${\to}$ $PV$ decays,
         \href{https://doi.org/10.1016/j.nuclphysb.2003.09.026}
              {Nucl. Phys. B 675, 333 (2003).}
     \bibitem{NuclPhysB.832.109}
         M. Beneke, T. Huber, X. Li,
         NNLO vertex corrections to non-leptonic $B$ decays: tree amplitudes,
         \href{https://doi.org/10.1016/j.nuclphysb.2010.02.002}
              {Nucl. Phys. B 832, 109 (2010).}
     \bibitem{PLB.748.422}
         C. Qiao, R. Zhu, X. Wu, S. Brodsky,
         A possible solution to the $B$ ${\to}$ ${\pi}{\pi}$ puzzle using the principle of maximum conformality,
         \href{https://dx.doi.org/10.1016/j.physletb.2015.07.044}
              {Phys. Lett. B 748, 422  (2015).}
     \bibitem{PhysRevD.80.114008}
         H. Cheng, C. Chua,
         Revisiting charmless hadronic $B_{u,d}$ decays in QCD factorization,
         \href{https://doi.org/10.1103/PhysRevD.80.114008}
              {Phys. Rev. D 80, 114008 (2009).}
     \bibitem{PhysRevD.90.054019}
         Q. Chang, J. Sun, Y. Yang, X. Li,
         Spectator scattering and annihilation contributions as a solution
         to the ${\pi}K$ and ${\pi}{\pi}$ puzzles within QCD factorization approach,
         \href{https://doi.org/10.1103/PhysRevD.90.054019}
              {Phys. Rev. D 90, 054019 (2014).}

     \bibitem{PhysRevD.74.014003}
         A. Williamson, J. Zupan,
         Two body $B$ decays with isosinglet final states in soft collinear effective theory,
         \href{https://doi.org/10.1103/PhysRevD.74.014003}
              {Phys. Rev. D 74, 014003 (2006);} Erratum,
         \href{https://doi.org/10.1103/PhysRevD.74.039901}
              {Phys. Rev. D 74, 039901 (2006).}

     \bibitem{PhysRevD.90.014029}
         Y. Zhang, X. Liu, Y. Fan, S. Cheng, Z. Xiao,
         $B$ ${\to}$ ${\pi}{\pi}$ decays and effects of the next-to-leading order contributions,
         \href{https://doi.org/10.1103/PhysRevD.90.014029}
              {Phys. Rev. D 90, 014029 (2014).}
     \bibitem{PhysRevD.91.114019}
         X. Liu, H. Li, Z. Xiao,
         Transverse-momentum-dependent wave functions with Glauber gluons
         in $B$ ${\to}$ ${\pi}{\pi}$, ${\rho}{\rho}$ decays,
         \href{https://doi.org/10.1103/PhysRevD.91.114019}
              {Phys. Rev. D 91, 114019 (2015).}
     \bibitem{PhysRevD.103.056006}
         Y. Yang, L. Lang, X. Zhao, J. Huang, J. Sun,
         Reinvestigating the $B$ ${\to}$ $PP$ decays by including the contributions from ${\phi}_{B2}$,
         \href{https://doi.org/10.1103/PhysRevD.103.056006}
              {Phys. Rev. D 103, 056006 (2021).}
     \bibitem{ChinPhysC.46.123103}
         J. Chai, S. Cheng, Y. Ju, D. Yan, C. L\"{u}, Z. Xiao,
         Charmless two-body $B$ meson decays in the perturbative QCD factorization approach,
         \href{https://doi.org/10.1088/1674-1137/ac88bd}
              {Chin. Phys. C 46, 123103 (2022).}
     \bibitem{PhysRevD.107.013004}
         S. L\"{u}, M. Yang,
         Possible solution of the puzzle for the branching ratio and $CP$
         violation in $B$ ${\to}$ ${\pi}{\pi}$ decays with a modified
         perturbative QCD approach,
         \href{https://doi.org/10.1103/PhysRevD.107.013004}
              {Phys. Rev. D 107, 013004 (2023).}
     \bibitem{PhysRevD.110.056025}
         S. L\"{u}, R. Wang, M. Yang,
         Study of $B$ ${\to}$ $PP$ decays in the modified perturbative QCD approach,
         \href{https://doi.org/10.1103/PhysRevD.110.056025}
              {Phys. Rev. D 110, 056025 (2024).}
     \bibitem{PhysRevD.108.013003}
         R. Wang, M. Yang,
         Branching ratio and $CP$ violation of $B$ ${\to}$ $K{\pi}$ decays
         in a modified perturbative QCD approach,
         \href{https://doi.org/10.1103/PhysRevD.108.013003}
              {Phys. Rev. D 108, 013003 (2023).}
     \bibitem{PhysRevD.111.036007}
         Y. Gui, M. Yang,
         Study of $B$ ${\to}$ ${\rho}{\eta}$, ${\rho}{\eta}^{\prime}$ decays in the modified perturbative QCD approach,
         \href{https://doi.org/10.1103/PhysRevD.111.036007}
              {Phys. Rev. D 111, 036007 (2025).}
     \bibitem{EPJC.85.146}
         R. Wang, M. Yang,
         Revisiting $B$ ${\to}$ $K^{\ast}{\pi}$ and $B$ ${\to}$ $K{\rho}$ decays using the modified perturbative QCD approach,
         \href{https://doi.org/10.1140/epjc/s10052-025-13840-8}
              {Eur. Phys. J. C 85, 146 (2025).}
     \bibitem{EPJC.85.821}
         S. L\"{u}, M. Yang,
         Study of $B$ ${\to}$ ${\pi}{\rho}$, ${\pi}{\omega}$ decays in the modified perturbative QCD approach,
         \href{https://doi.org/10.1140/epjc/s10052-025-14527-w}
              {Eur. Phys. J. C 85, 821 (2025).}
     \bibitem{JPG.52.105001}
         R. Wang, M. Yang,
         Branching fraction, $CP$ asymmetry, and polarization in $B$ ${\to}$ ${\rho}{\rho}$
         decays with the modified perturbative QCD approach,
         \href{https://doi.org/10.1088/1361-6471/ae0659}
              {J. Phys. G 52, 105001 (2025).}

     \bibitem{PLB.52.351}
         G. Altarelli, L. Maiani,
         Octet enhancement of nonleptonic weak interactions in asymptotically free gauge theories,
         \href{https://doi.org/10.1016/0370-2693(74)90060-4}
              {Phys. Lett. B 52, 351 (1974).}
     \bibitem{PLB.86.343}
         H. Fritzsch,
         How to discover the $B$ mesons,
         \href{https://doi.org/10.1016/0370-2693(79)90853-0}
              {Phys. Lett. B 86, 343 (1979).}
     \bibitem{PLB.90.431}
         N. Deshpande, M. Gronau, D. Sutherland,
         Final state gluon effects in charmed meson decays,
         \href{https://doi.org/10.1016/0370-2693(80)90967-3}
              {Phys. Lett. B 90, 431 (1980).}
     \bibitem{PLB.93.125}
         Y. Igarashi, S. Kitakado, M. Kuroda,
         Soft hadron interaction effects in charmed meson decays,
         \href{https://doi.org/10.1016/0370-2693(80)90109-4}
              {Phys. Lett. B 93, 125 (1980).}
     \bibitem{NPB.171.413}
         H. Fritzsch, P. Minkowski,
         Tests of the quark annihilation hypothesis for $D$ and $F$ decays,
         \href{https://doi.org/10.1016/0550-3213(80)90378-8}
              {Nucl. Phys. B 171, 413 (1980).}
     \bibitem{NPB.181.74}
         J. Finjord,
         The magnitude of annihilation and penguin contributions to charm meson decays,
         \href{https://doi.org/10.1016/0550-3213(81)90507-1}
              {Nucl. Phys. B 181, 74 (1981);} Erratum:
         \href{https://doi.org/10.1016/0550-3213(82)90342-X}
              {Nucl. Phys. B 198, 544 (1982).}
     \bibitem{NPB.183.367}
         M. Gronau, D. Sutherland,
         Soft gluonic effects in non-leptonic decays,
         \href{https://doi.org/10.1016/0550-3213(81)90139-5}
              {Nucl. Phys. B 183, 367 (1981).}
     \bibitem{PLB.100.420}
         Q. Ho-Kim, H. Lee,
         Gluon effects in charmed meson decays,
         \href{https://doi.org/10.1016/0370-2693(81)90150-7}
              {Phys. Lett. B 100, 420  (1981).}
     \bibitem{PLB.109.227}
         J. Finjord,
         Decays of charmed mesons via color octet decays of the $c$-quark,
         \href{https://doi.org/10.1016/0370-2693(82)90759-6}
              {Phys. Lett. B 109, 227  (1982).}
     \bibitem{PhysRevD.25.178}
         Q. Ho-Kim, H. Lee,
         Decays of charmed mesons,
         \href{https://doi.org/10.1103/PhysRevD.25.178}
              {Phys. Rev. D 25, 178 (1982).}
     \bibitem{NPB.268.16}
         A. Buras, J. G\'{e}rard, R. R\"{u}ckl,
         $1/N_{c}$ expansion for exclusive and inclusive charm decays,
         \href{https://doi.org/10.1016/0550-3213(86)90200-2}
              {Nucl. Phys. B 268, 16 (1986).}
     \bibitem{ZPC.32.237}
         H. Cheng,
         Vacuum insertion and nonperturbative effects for exclusive nonleptonic decays of charmed mesons and kaons,
         \href{https://doi.org/10.1007/BF01552501}
              {Z. Phys. C 32, 237 (1986).}
     \bibitem{IJMPA.4.495}
         H. Cheng,
         Status of the ${\Delta}I$ $=$ $\frac{1}{2}$ rule in kaon decay,
         \href{https://doi.org/10.1142/S0217751X89000261}
              {Int. J. Mod. Phys. A 4, 495 (1989).}
     \bibitem{PLB.318.549}
         A. Deandrea, N. Bartolomeo, R. Gatto, G. Nardulli,
         Two-body nonleptonic decays of $B$ and $B_{s}$ mesons,
         \href{https://doi.org/10.1016/0370-2693(93)91554-Z}
              {Phys. Lett. B 318, 549 (1993).}
     \bibitem{NPB.389.534}
         B. Blok, M. Shifman,
         Nonfactorizable amplitudes in weak nonleptonic decays of heavy mesons,
         \href{https://doi.org/10.1016/0550-3213(93)90330-R}
              {Nucl. Phys. B 389, 534 (1993).}
     \bibitem{NPB.399.441}
         B. Blok, M. Shifman,
         The rule of discarding $1/N_{c}$ in inclusive weak decays (I),
         \href{https://doi.org/10.1016/0550-3213(93)90504-I}
              {Nucl. Phys. B 399, 441 (1993).}
     \bibitem{PLB.335.428}
         H. Cheng,
         Nonfactorizable contributions to nonleptonic weak decays of heavy mesons,
         \href{https://doi.org/10.1016/0370-2693(94)90374-3}
              {Phys. Lett. B 335, 428 (1994).}
     \bibitem{PLB.349.548}
         I. Halperin,
         Soft gluon suppression of $1/N_{c}$ contributions in color
         suppressed heavy meson decays,
         \href{https://doi.org/10.1016/0370-2693(95)00264-L}
              {Phys. Lett. B 349, 548 (1995).}
     \bibitem{PhysRevD.51.3518}
         J. Soares,
         ``Nonfactorizable'' terms in hadronic $B$-meson weak decays,
         \href{https://doi.org/10.1103/PhysRevD.51.3518}
              {Phys. Rev. D 51, 3518 (1995).}
     \bibitem{PhysRevD.51.6259}
         H. Cheng, B. Tseng,
         Extraction of $a_{1}$ and $a_{2}$ from $B$ ${\to}$ ${\psi}K(K^{\ast})$,
         $D(D^{\ast}){\pi}({\rho})$ decays,
         \href{https://doi.org/10.1103/PhysRevD.51.6259}
              {Phys. Rev. D 51, 6259 (1995).}
     \bibitem{PhysRevD.53.2506}
         A. Kamal, A. Santra, T. Uppal, R. Verma,
         Nonfactorization in hadronic two-body Cabibbo favored decays of $D^{0}$ and $D^{+}$,
         \href{https://doi.org/10.1103/PhysRevD.53.2506}
              {Phys. Rev. D 53, 2506 (1996).}
     \bibitem{ZPC.69.253}
         R. Verma ,
         Searching a systematics for nonfactorizable contributions to hadronic decays of $D^{0}$ and $D^{+}$ mesons,
         \href{https://doi.org/10.1007/BF02907405}
              {Z. Phys. C 69, 253 (1996).}
     \bibitem{ZPC.69.647}
         H. Cheng,
         Hadronic weak decays of heavy mesons and nonfactorization,
         \href{https://doi.org/10.1007/s002880050068}
              {Z. Phys. C 69, 647 (1996).}
     \bibitem{ZPC.71.101}
         A. Kamal, A. Santra,
         Nonfactorization and the decays $D_{s}^{+}$ ${\to}$ ${\phi}{\pi}^{+}$,
         ${\phi}{\rho}^{+}$, and ${\phi}{\ell}^{+}{\nu}_{\ell}$,
         \href{https://doi.org/10.1007/s002880050152}
              {Z. Phys. C 71, 101 (1996).}
     \bibitem{ZPC.72.91}
         A. Kamal, A. Santra,
         Nonfactorization and color-suppressed $B$ ${\to}$ ${\psi}({\psi}^{\prime})$ $+$ $K(K^{\ast})$ decays,
         \href{https://doi.org/10.1007/s002880050227}
              {Z. Phys. C 72, 91 (1996).}
     \bibitem{PLB.393.149}
         J. Donoghue, A. Petrov,
         Factorization in nonleptonic decays of heavy mesons,
         \href{https://doi.org/10.1016/S0370-2693(96)01607-3}
              {Phys. Lett. B 393, 149 (1997).}
     \bibitem{PLB.395.345}
         H. Cheng,
         Can $B$ ${\to}$ $J/{\psi}K(K^{\ast})$ decays be described by factorization?,
         \href{https://doi.org/10.1016/S0370-2693(97)00083-X}
              {Phys. Lett. B 395, 345 (1997).}
     \bibitem{ZPC.75.651}
         A. Katoch, K. Sharma,
         Nonfactorizable contributions to hadronic decays of charm mesons in the Cabibbo-suppressed modes,
         \href{https://doi.org/10.1007/s002880050510}
              {Z. Phys. C 75, 651 (1997).}
     \bibitem{ZPC.76.311}
         K. Sharma, A. Katoch, R. Verma,
         Nonfactorizable contributions to charm meson ($D$ ${\to}$ $PP$) decays,
         \href{https://doi.org/10.1007/s002880050555}
              {Z. Phys. C 76, 311 (1997).}
     \bibitem{JPG.23.807}
         A. Katoch, K. Sharma, R. Verma,
         Isospin analysis of non-factorizable contributions to hadronic decays of charm mesons,
         \href{https://doi.org/10.1088/0954-3899/23/7/006}
              {J. Phys. G 23, 807 (1997).}
     \bibitem{EPJC.4.669}
         F. Al-Shamali, A. Kamal,
         Nonfactorization and final state interactions in $(B,B_{s})$ ${\to}$ ${\psi}P$ and ${\psi}V$ decays,
         \href{https://doi.org/10.1007/s100529800798}
              {Eur. Phys. J. C 4, 669 (1998).}
     \bibitem{PhysRevD.58.094005}
         H. Cheng, B. Tseng,
         Nonfactorizable effects in spectator and penguin amplitudes of hadronic charmless $B$ decays,
         \href{https://doi.org/10.1103/PhysRevD.58.094005}
              {Phys. Rev. D 58, 094005 (1998).}
     \bibitem{Adv.Ser.Direct.HEP.15.294}
         M. Neubert, B. Stech,
         Nonleptonic weak decays of $B$ mesons,
         \href{https://doi.org/10.1142/9789812812667_0004}
              {Adv. Ser. Direct. High Energy Phys. 15, 294 (1998).}
     \bibitem{PhysRevD.59.054020}
         F. Al-Shamali, A. Kamal,
         Nonfactorization in Cabibbo favored $B$ decays,
         \href{https://doi.org/10.1103/PhysRevD.59.054020}
              {Phys. Rev. D 59, 054020 (1999).}
     \bibitem{MPLA.14.1841}
         K. Sharma, A. Katoch,
         An isospin analysis of nonfactorizable contributions to hadronic decays of bottom mesons,
         \href{https://doi.org/10.1142/S0217732399001929}
              {Mod. Phys. Lett. A 14, 1841 (1999).}
     \bibitem{PhysRevD.59.074003}
         Y. Chen, H. Cheng, B. Tseng,
         Charmless hadronic two-body decays of $B_{s}$ mesons,
         \href{https://doi.org/10.1103/PhysRevD.59.074003}
              {Phys. Rev. D 59, 074003 (1999).}
      \bibitem{PhysRevD.59.092004}
         H. Cheng, K. Yang,
         Updated analysis of $a_{1}$ and $a_{2}$ in hadronic two-body decays of $B$ mesons,
         \href{https://doi.org/10.1103/PhysRevD.59.092004}
              {Phys. Rev. D 59, 092004 (1999).}
     \bibitem{PhysRevD.60.114019}
         F. Al-Shamali, A. Kamal,
         Factorization and nonfactorization in $B$ decays,
         \href{https://doi.org/10.1103/PhysRevD.60.114019}
              {Phys. Rev. D 60, 114019 (1999).}
     \bibitem{NPB.605.558}
         A. Khodjamirian,
         $B$ ${\to}$ ${\pi}{\pi}$ decay in QCD,
         \href{https://doi.org/10.1016/S0550-3213(01)00194-8}
              {Nucl. Phys. B 605, 558 (2001).}
     \bibitem{PLB.571.75}
         A. Khodjamirian, Th. Mannel, B. Meli\'{c},
         QCD light-cone sum rule estimate of charming penguin contributions in $B$ ${\to}$ ${\pi}{\pi}$,
         \href{https://doi.org/10.1016/j.physletb.2003.08.012}
              {Phys. Lett. B 571, 75 (2003).}
     \bibitem{PhysRevD.68.114009}
         S. Mantry, D. Pirjol, I. Stewart,
         Strong phases and factorization for color suppressed decays,
         \href{https://doi.org/10.1103/PhysRevD.68.114009}
              {Phys. Rev. D 68, 114009 (2003).}
     \bibitem{PhysRevD.72.094012}
         A. Khodjamirian, Th. Mannel, M. Melcher, B. Meli\'{c},
         Annihilation effects in $B$ ${\to}$ ${\pi}{\pi}$ from QCD light-cone sum rules,
         \href{https://doi.org/10.1103/PhysRevD.72.094012}
              {Phys. Rev. D 72, 094012 (2005).}
     \bibitem{JPG.34.637}
         R. Dhir, R. Verma,
         Nonfactorizable contributions to hadronic decays of $D$ mesons,
         \href{https://doi.org/10.1088/0954-3899/34/4/003}
              {J. Phys. G 34, 637  (2007).}
     \bibitem{PLB.675.196}
         X. Wu, B. Zhang, H. Li, X. Liu, B. Liu, J. Li, Y. Guo,
         Non-factorization contributions in $D$ ${\to}$ ${\pi}K$, $KK$ decay,
         \href{https://doi.org/10.1016/j.physletb.2009.04.006}
              {Phys. Lett. B 675, 196 (2009).}
     \bibitem{CPC.46.073105}
         M. Kaur, S. Singh, R. Verma,
         Searching systematics for nonfactorizable contributions to
         $B^{-}$ and $\overline{B}^{0}$ hadronic decays,
         \href{https://doi.org/10.1088/1674-1137/ac600b}
              {Chin. Phys. C 46, 073105 (2022).}
     \bibitem{JHEP.2023.10.180}
         M. Piscopo, A. Rusov,
         Non-factorisable effects in the decays $\overline{B}_{s}^{0}$ ${\to}$
         $D_{s}^{+}{\pi}^{-}$ and $\overline{B}^{0}$ ${\to}$ $D^{+}K^{-}$ from LCSR,
         \href{https://doi.org/10.1007/JHEP10(2023)180}
              {JHEP 10, 180 (2023).}
     \bibitem{CPC.49.023105}
         M. Kaur, S. Singh, R. Verma,
         $p$-wave mesons emitting weak decays of bottom mesons,
         \href{https://doi.org/10.1088/1674-1137/ad9893}
              {Chin. Phys. C 49, 023105 (2025).}

     \bibitem{RevModPhys.68.1125}
         G. Buchalla, A. Buras, M. Lautenbacher,
         Weak decays beyond leading logarithms,
         \href{https://doi.org/10.1103/RevModPhys.68.1125}
              {Rev. Mod. Phys. 68, 1125, (1996).}

     \bibitem{PRL.51.1945}
         L. Wolfenstein,
         Parametrization of the Kobayashi-Maskawa Matrix,
         \href{https://doi.org/10.1103/PhysRevLett.51.1945}
              {Phys. Rev. Lett. 51, 1945 (1983).}
     \bibitem{PRD.50.3433}
         A. Buras, M. Lautenbacher, G. Ostermaier,
         Waiting for the top quark mass, $K^{+}$ ${\to}$ ${\pi}^{+}{\nu}\bar{\nu}$,
         $B_{s}^{0}$-$\overline{B}_{s}^{0}$ mixing, and $CP$ asymmetries in $B$ decays,
         \href{https://doi.org/10.1103/PhysRevD.50.3433}
              {Phys. Rev. D 50, 3433 (1994).}

     \bibitem{JHEP.0605.004}
         P. Ball, V. Braun, A. Lenz,
         Higher-twist distribution amplitudes of the $K$ meson in QCD,
         \href{https://doi.org/10.1088/1126-6708/2006/05/004}
         {JHEP 05, 004 (2006).}

     \bibitem{NPB.22.448}
         O. Haan, B. Stech,
         Violation of the ${\Delta}I$ $=$ $1/2$ rule in non-leptonic decays,
         \href{https://doi.org/10.1016/0550-3213(70)90420-7}
              {Nucl. Phys. B 22, 448 (1970).}
     \bibitem{RevModPhys.47.277}
         M. Gaillard, B. Lee, J. Rosner, Search for charm,
         \href{https://doi.org/10.1103/RevModPhys.47.277}
              {Rev. Mod. Phys. 47, 277 (1975).}
     \bibitem{NPB.100.313}
         J. Ellis, M. Gaillard, D. Nanopoulos,
         On the weak decays of high mass hadrons,
         \href{https://doi.org/10.1016/0550-3213(75)90621-5}
              {Nucl. Phys. B 100, 313 (1975);} Erratum:
         \href{https://doi.org/10.1016/0550-3213(76)90118-8}
              {Nucl. Phys. B 104, 547 (1976).}
     \bibitem{NPB.133.315}
         D. Fakirov, B. Stech,
         $F$ and $D$ decays,
         \href{https://doi.org/10.1016/0550-3213(78)90306-1}
              {Nucl. Phys. B 133, 315 (1978).}
     \bibitem{NPB.201.315}
         V. Chernyak, A. Zhitnitsky,
         Exclusive decays of heavy mesons,
         \href{https://doi.org/10.1016/0550-3213(82)90445-X}
              {Nucl. Phys. B 201, 492 (1982);}  Erratum:
         \href{https://doi.org/10.1016/0550-3213(83)90251-1}
              {Nucl. Phys. B 214, 547 (1983).}
     \bibitem{NPB.264.371}
         A. Buras, J. G\'{e}rard,
         $1/N$ expansion for kaons,
         \href{https://doi.org/10.1016/0550-3213(86)90489-X}
              {Nucl. Phys. B 264, 371 (1986).}
     \bibitem{ZPC.34.103}
         M. Bauer, B. Stech, M. Wirbel,
         Exclusive nonleptonic decays of $D$-, $D_{s}$-, and $B$ mesons,
         \href{https://doi.org/10.1007/BF01561122}
              {Z. Phys. C 34, 103 (1987).}

     \bibitem{NPB.PS.11.325}
         J. Bjorken, Topics in $B$-physics,
         \href{https://doi.org/10.1016/0920-5632(89)90019-4}
              {Nucl. Phys. B Proc. Suppl. 11, 325 (1989).}

     \bibitem{ZPC.29.637}
         M. Wirbel, B. Stech,
         M Bauer, Exclusive semileptonic decays of heavy mesons,
         \href{https://doi.org/10.1007/BF01560299}
              {Z. Phys. C 29, 637 (1985).}

     \bibitem{PLB.90.455}
         H. Fritzsch, P. Minkowski,
         The puzzle of nonleptonic decays and its resolution,
         \href{https://doi.org/10.1016/0370-2693(80)90973-9}
              {Phys. Lett. B 90, 455 (1980).}

     \bibitem{PRD.22.2157}
         G. Lepage, S. Brodsky,
         Exclusive processes in perturbative quantum chromodynamics,
         \href{https://doi.org/10.1103/PhysRevD.22.2157}
              {Phys. Rev. D 22, 2157 (1980).}

     \bibitem{JHEP.2020.04.055}
         G. Bell, M. Beneke, T. Huber, X. Li,
         Two-loop non-leptonic penguin amplitude in QCD factorization,
         \href{https://doi.org/10.1007/JHEP04(2020)055}
              {JHEP 04, 055 (2020).}

     \bibitem{PLB.740.56}
         Q. Chang, J. Sun, Y. Yang, X. Li,
         A combined fit on the annihilation corrections in $B_{u,d,s}$
         ${\to}$ $PP$ decays within QCDF,
         \href{https://doi.org/10.1016/j.physletb.2014.11.027}
              {Phys. Lett. B 740, 56 (2015).}
     \bibitem{PLB.743.444}
         J. Sun, Q. Chang, X. Hu, Y. Yang,
         Constraints on hard spectator scattering and annihilation
         corrections in $B_{u,d}$ ${\to}$ $PV$ decays within QCD factorization,
         \href{https://doi.org/10.1016/j.physletb.2015.03.001}
              {Phys. Lett. B 743, 444 (2015).}
     \bibitem{PhysRevD.91.074026}
         Q. Chang, X. Hu, J. Sun, Y. Yang,
         Probing spectator scattering and annihilation corrections in $B_{s}$ ${\to}$ $PV$ decays,
         \href{https://doi.org/10.1103/PhysRevD.91.074026}
              {Phys. Rev. D 91, 074026 (2015).}
     \bibitem{JPG.43.105004}
         Q. Chang, X. Li, J. Sun, Y. Yang,
         Fits of weak annihilation and hard spectator scattering corrections in $B_{u,d}$ ${\to}$ $VV$ decays,
         \href{https://doi.org/10.1088/0954-3899/43/10/105004}
              {J. Phys. G 43, 105004 (2016).}
     \bibitem{PhysRevD.88.014043}
         K. Wang, G. Zhu,
         Flavor dependence of annihilation parameters in QCD factorization,
         \href{https://doi.org/10.1103/PhysRevD.88.014043}
              {Phys. Rev. D 88, 014043 (2013).}

     \bibitem{PhysRevD.106.054502}
         B. Colquhoun, S. Hashimoto, T. Kaneko, J. Koponen (JLQCD Collaboration),
         Form factors of $B$ ${\to}$ ${\pi}{\ell}{\nu}$ and a determination
         of ${\vert}V_{ub}{\vert}$ with M\"{o}bius domain-wall fermions,
         \href{https://doi.org/10.1103/PhysRevD.106.054502}
              {Phys. Rev. D 106, 054502 (2022).}
     \bibitem{JHEP.2023.03.140}
         B. Cui, Y. Huang, Y. Shen, C. Wang, Y. Wang,
         Precision calculations of $B_{d,s}$ ${\to}$ ${\pi}$, $K$ decay
         form factors in soft-collinear effective theory,
         \href{https://doi.org/10.1007/JHEP03(2023)140}
              {JHEP 03, 140 (2023).}
     \end{thebibliography}
     \end{document}